\documentclass[a4paper]{article}
\usepackage[dvips]{graphicx}
\usepackage{verbatim}
\usepackage{amssymb}
\usepackage{amsmath}
\usepackage{amsbsy}
\usepackage{amsfonts}
\usepackage{amsthm}
\usepackage{amsxtra}
\usepackage{t1enc}
\usepackage{bbm}
\usepackage{mathrsfs}
\usepackage{ifthen}

\begin{document}

% THEOREM TYPES
% Define a theorem style named "kalle":
\newtheoremstyle{thm-style-kalle}
{7pt}      % Space above
{7pt}      % Space below
{\itshape} % Body font
{}         % Indent amount (empty = no indent, \parindent = para indent)
{\scshape} % Thm head font
{.}        % Punctuation after thm head
{.5em}     % Space after thm head: " " = normal interword space; 
           % \newline = linebreak
{}         % Thm head spec (can be left empty, meaning `normal')

% This style will be used until other style is given:
\theoremstyle{thm-style-kalle}
    % Number w.r.t. the sections
    \newtheorem{theorem}{Theorem}[section]
    % For the rest the numbering is like in "theorem":
    \newtheorem{proposition}[theorem]{Proposition}
    \newtheorem{corollary}[theorem]{Corollary}
    \newtheorem{lemma}[theorem]{Lemma}
%\theorembodyfont{\rmfamily} 
    \newtheorem{definition}[theorem]{Definition}
    \newtheorem{example}{Example}[section]
    \newtheorem{remark}{Remark}[section]

\newenvironment{theoremn}[1] % Theorem with a NAME
    {\begin{theorem} \emph{(#1)}}{\end{theorem}}
\newenvironment{theoremcn}[2] % Theorem with a CITATION and a NAME
    {\begin{theorem} \cite{#1} \emph{(#2)}}{\end{theorem}}
\newenvironment{theoremc}[1] % Theorem with a CITATION
    {\begin{theorem} \cite{#1}}{\end{theorem}}
\newenvironment{propositionn}[1] % Proposition with a NAME
    {\begin{proposition} \emph{(#1)}}{\end{proposition}}
\newenvironment{propositioncn}[2] % Proposition with a CITATION and a NAME
    {\begin{proposition} \cite{#1} \emph{(#2)}}{\end{proposition}}
\newenvironment{definitionn}[1] % Definition with a NAME
    {\begin{definition} \emph{(#1)}}{\end{definition}}
\newenvironment{Proof}[1][Proof]{\begin{proof}[\sc{#1}]}{\end{proof}}

\newcommand{\Remark} {\emph{Remark:  }}
\newcommand{\QED} {$\quad \square$ \\}
%\newcommand{\sQED} {$\boxtimes$ \\}

% GREEK LETTERS
\newcommand{\la} {\lambda}
\newcommand{\de} {\delta}
\newcommand{\be} {\beta}
\newcommand{\si} {\sigma}
\newcommand{\om} {\omega}
\newcommand{\te} {\theta}
\newcommand{\vphi} {\varphi}
\newcommand{\eps} {\varepsilon}
\newcommand{\Om} {\Omega}
\newcommand{\La} {\Lambda}
\newcommand{\Th} {\Theta}

% CAPITAL LETTERS
\newcommand{\bH} {\mathbb{H}}  % upper half plane
\newcommand{\bC} {\mathbb{C}}  % complex plane
\newcommand{\bR} {\mathbb{R}}  % real line
\newcommand{\bN} {\mathbb{N}}  % natural numbers 1,2,
\newcommand{\bZ} {\mathbb{Z}}  % integers
\newcommand{\bQ} {\mathbb{Q}}  % rational numbers
\newcommand{\bD} {\mathbb{D}}  % unit disc
\newcommand{\sD} {\mathcal{D}} % 
\newcommand{\sH} {\mathcal{H}} % 
\newcommand{\sL} {\mathcal{L}} %
\newcommand{\sI} {\mathcal{I}} %
\newcommand{\sA} {\mathcal{A}} %
\newcommand{\sV} {\mathcal{V}} % 
\newcommand{\sP} {\mathcal{P}} %
\newcommand{\sS} {\mathcal{S}} %
\newcommand{\sR} {\mathcal{R}} %
\newcommand{\sF} {\mathcal{F}} %
\newcommand{\sM} {\mathcal{M}} %
\newcommand{\sT} {\mathcal{T}} %
\newcommand{\sK} {\mathcal{K}} %
\newcommand{\sU} {\mathcal{U}} %
\newcommand{\sQ} {\mathcal{Q}} %

% PROBABILITY SYMBOLS
\newcommand{\PR} {\mathsf{P}}
\newcommand{\EX} {\mathsf{E}}
\newcommand{\lE} {\Big\langle}
\newcommand{\rE} {\Big\rangle}
\newcommand{\qvl} {\langle}
\newcommand{\qvr} {\rangle}

% MISCELLANEOUS LETTERS
\newcommand{\ud} {\mathrm{d}}
\newcommand{\bn} {\mathbf{n}}
\newcommand{\bm} {\mathbf{m}}
\newcommand{\bl} {\mathbf{l}}
\newcommand{\bg} {\mathbf{g}}
\newcommand{\bh} {\mathbf{h}}
\newcommand{\bv} {\mathbf{v}}
\newcommand{\bu} {\mathbf{u}}
\newcommand{\bfs} {\mathbf{f}}
\newcommand{\bs} {\mathbf{s}}
\newcommand{\bc} {\mathbf{c}}
\newcommand{\bte} {\boldsymbol{\theta}}
\newcommand{\bphi} {\boldsymbol{\phi}}
\newcommand{\bsi} {\boldsymbol{\sigma}}
\newcommand{\bta} {\boldsymbol{\tau}}
\newcommand{\bet} {\boldsymbol{\eta}}
\newcommand{\bel} {\boldsymbol{\ell}}
\newcommand{\bro} {\boldsymbol{\rho}}
\newcommand{\bde} {\boldsymbol{\delta}}
\newcommand{\bzero} {\boldsymbol{0}}
\newcommand{\bone} {\boldsymbol{1}}

% RELATIONS
\newcommand{\conn} {\leftrightarrow}
\newcommand{\nconn} {\nleftrightarrow}

% OPERATIONS
%\newcommand{\mod} {\mathrm{ mod } \,}
\newcommand{\pder}[1] {\frac{\partial}{\partial #1}}
\newcommand{\ppder}[1] {\frac{\partial^2}{\partial #1^2}}
\newcommand{\Order} {\mathcal{O}}
\newcommand{\trans} {^{\mathsf{T}}}
\newcommand{\tp} {^{\mathsf{T}}}
\newcommand{\til} {\Tilde}
\newcommand{\wtil} {\widetilde}
\newcommand{\lrot} {\nabla \times}
\newcommand{\lgrad} {\nabla}
\newcommand{\ldiv} {\nabla \cdot}
\newcommand{\lapl} {\triangle}
\newcommand{\fprod} {\boxtimes}
\newcommand{\lconj} {\overline \nabla}
\newcommand{\unit} {\mathbf{1}}
\newcommand{\Imag} {\mathrm{Im } \; }
\newcommand{\Kern} {\mathrm{Ker } \; }
\newcommand{\End} {\mathrm{End } \; }
\newcommand{\Res}[1] {\mathrm{Res}_{#1} \; }
\newcommand{\oires}[2] {\frac{1}{2 \pi i}
    \oint_\infty \left( {#2} \right) \, \ud {#1} }
\newcommand{\oiresb}[2] {\frac{1}{2 \pi i}
    \oint_\infty \big( {#2} \big) \, \ud {#1} }
\newcommand{\oiresB}[2] {\frac{1}{2 \pi i}
    \oint_\infty \Big( {#2} \Big) \, \ud {#1} }
\newcommand{\oiresn}[2] {\frac{1}{2 \pi i}
    \oint_\infty {#2} \; \ud {#1} }
\newcommand{\cconj} {\overline}
\newcommand{\bbar} {\underline}
\newcommand{\tbar} {\overline}
\newcommand{\id} {\mathrm{id}}
\newcommand{\sign} {\mathrm{sign } \; }
\newcommand{\spn} {\mathrm{span } \; }
\newcommand{\dmn} {\mathrm{dim } \; }
\newcommand{\ann} {\mathbf{a}}
\newcommand{\cre} {\mathbf{a}^\dag}
\newcommand{\num} {\mathbf{N}}
\newcommand{\no} {\mathbf{:}}
\newcommand{\diam} {\mathrm{ diam}}
\newcommand{\conv} {*}
\newcommand{\ldual} {\big<}
\newcommand{\rdual} {\big>}
\newcommand{\bra} {\big<}
\newcommand{\ket} {\big>}
\newcommand{\im} {\Im \textrm{m }}
\newcommand{\re} {\Re \textrm{e }}
\newcommand{\ncdots} {\cdot \! \cdot \! \cdot}

% TOPOLOGY
\newcommand{\cl} {\overline}
\newcommand{\bdry} {\partial}

% LIE ALGEBRA
\newcommand{\ag} {\hat{\mathfrak{g}}}
\newcommand{\ssg} {\mathfrak{g}}
\newcommand{\vir} {\mathfrak{vir}}
\newcommand{\sn} {\mathfrak{n}}
\newcommand{\irhwm} {\mathcal{H}}
\newcommand{\Verma} {\mathcal{V}}
\newcommand{\Stagg} {\mathcal{S}}
\newcommand{\vac} {\Psi}
\newcommand{\hwvV} {\eta}
\newcommand{\hwvI} {\omega}
\newcommand{\ghwv} {w}
\newcommand{\zero} {\omega}
\newcommand{\ind} {\mathbf{1}}
\newcommand{\isom} {\cong}
\newcommand{\uea} {\mathcal{U}}

% ABBREVIATIONS
\newcommand{\half} {\frac{1}{2}}
\newcommand{\const} {\mathrm{const.}}
\newcommand{\expand}[2] {|#1| \; \mathrm{ #2}}
\newcommand{\Expand}[2] {|#1| > |#2|}
\newcommand{\fus} {\mathrm{f}}
\newcommand{\poly} {\mathrm{polyn.}}
\newcommand{\checked} {$\quad$ {\small \scshape ok.}}
\newcommand{\general} {$\quad$ {\small \scshape gen.}}
\newcommand{\Arev} {{A^\downarrow}}
\newcommand{\lft} {\mathrm{L}}
\newcommand{\rgt} {\mathrm{R}}

\title{SLE local martingales in logarithmic representations}
%\author{Kalle Kyt\"ol\"a}
\author{}
\date{}
\maketitle

\centerline{Kalle Kyt\"ol\"a}
\centerline{\small \texttt{kalle.kytola@math.unige.ch}}

\bigskip

\centerline{Section de mathématiques, Université de Genève}
\centerline{2-4 Rue du Lièvre, CP 64, 1211 Genève 4, Switzerland.}

\bigskip

\begin{abstract}
A space of local martingales of SLE type growth processes forms a
representation of Virasoro algebra, but apart from a few simplest
cases not much is known about this representation. The purpose of
this article is to exhibit examples of representations
where $L_0$ is not diagonalizable --- a phenomenon characteristic
of logarithmic conformal field theory. Furthermore, we observe that
the local martingales bear a close relation with the
fusion product of the boundary changing fields.

Our examples reproduce first of all many familiar logarithmic
representations at certain rational values of the central charge. In particular
we discuss the case of SLE${}_{\kappa=6}$ describing the exploration
path in critical percolation, and its relation with the question of
operator content of the appropriate conformal field theory of zero central
charge. In this case one encounters logarithms in a probabilistically
transparent way, through conditioning on a crossing event.
But we also observe that some quite natural SLE variants exhibit
logarithmic behavior at all values of $\kappa$, thus at all central
charges and not only at specific rational values.
\end{abstract}

%  *****************************************************************
%  *  INTRODUCTION  ************************************************
%  *****************************************************************

\section{Introduction}
\label{sec: intro}

Conformal invariance is without a doubt one of the most powerful concepts
in the research of two dimensional critical phenomena.
When the microscopic lengthscale provided
by a lattice is taken to zero in a model of statistical
mechanics at its critical point, the scaling limit is expected to be
conformally invariant in a suitable sense.
If the scaling limit is understood in terms of a quantum
field theory, one speaks of conformal field theory (CFT)
\cite{BPZ-1984, DMS-CFT}.
Applying conformal field theory to statistical mechanics has lead
to numerous great successes since 1980's --- some of the most
fundamental and best established results concern exact values of
critical exponents, correlation functions and
partition functions (in particular modular invariant ones on the torus).

The way the conformal invariance assumption is exploited in conformal field
theories is through the effect that infinitesimal local conformal
transformations have on correlation functions of local fields.
The (central extension of the) Lie algebra of these infinitesimal
transformations is the Virasoro algebra,
and it seems fair to say that the representation theory of Virasoro algebra
forms the basis of all conformal field theory. In particular,
conformal field theories built from a finite number of
irreducible highest weight representations,
the so called minimal models, were able to explain critical
exponents in many of the most important two dimensional models of
statistical physics.

It is nevertheless worth noting that in some applications
of conformal field theory, minimal models were found insufficient.
Among notable examples were percolation and polymers of different kinds
\cite{Saleur-conf_inv_polymers_and_percolation, SR-S_and_T_matrices,
Cardy-logarithmic_correlations}.
These cases often required something called logarithmic
conformal field theory (LCFT) --- the assumption that had to be relaxed from
representation theory point of view was that of diagonalizability of $L_0$,
the Virasoro generator corresponding physically to energy
\cite{Gurarie-logarithmic_operators}. 
%In hindsight, one can also argue that many of the models
%that required such generalizations deal with geometric objects, for
%example the polymers themselves or percolation clusters.
%It has indeed recently been suggested as a sort of a rule of thumb that
%answering questions of global geometric nature in CFTs, rather than just
%questions of correlations of local degrees of freedom at a distance,
%typically requires logarithmic conformal field theory
%\cite{SPR-geometric_exponents}.

Since almost a decade now, the conformal invariance assumption has been
exploited also in another, rather different way. The introduction
of Schramm-Loewner evolutions (SLE) \cite{Schramm-LERW_and_UST}
marked the starting point in the study of random curves
whose probability distribution was postulated to be conformally invariant.
Supplemented
with another natural assumption from statistical physics, the requirement
of conformal invariance of the probability distribution
is powerful enough to
identify the possible random curves up to one free parameter $\kappa > 0$
--- resulting in what is known as SLE${}_\kappa$.
More accurately, this is true when one chooses boundary conditions
conveniently, whereas more general boundary conditions require introducing
variants of SLEs. The approach has been very fruitful for making rigorous
connection with lattice models of statistical mechanics. It has been proven
in several cases that the law of an interface in a critical lattice model
converges to SLE as the lattice spacing is taken to zero.

When it comes to the necessary mathematical techniques, these new methods
differ significantly from conformal field theories. The very
definition of SLEs does use a complex analysis technique of Loewner
chains \cite{Loewner-schlicht}
essentially composing infinitesimal conformal maps
to describe the curve as a growth process. But to answer questions about the
random curves, calculations mostly involve It\^o's stochastic
analysis: a crucial step in solving an SLE problem is often that of finding
an appropriate martingale.

Despite the differences of the paradigms of conformal field theory
and Schramm-Loewner evolutions, there is an increasing amount of overlap.
The purpose of the present article is to continue building a bridge
between the two. The particular example of an interplay of SLE and CFT
concepts relevant to us
is the observation that local martingales of the SLE growth processes
carry a representation of the Virasoro algebra \cite{Kytola-local_mgales}.
We will study this representation in some detail, and in doing so
show the relevance to SLEs of yet other conformal field theory concepts:
the above mentioned logarithmic conformal field theories, and fusion
products.
Concretely, in Sections \ref{sec: representations}
and \ref{sec: kappa 6} we will examine explicitly for several variants
of SLE the structure of the Virasoro representation consisting
of local martingales. We will in
particular find examples where $L_0$ is not diagonalizable.
The results suggest a very close relationship between the local martingales
and the fusion product of the boundary condition changing fields of
the variant. More precisely, in all the examples we observe that the
contragredient of the fusion product can be realized as local martingales.

In parallel with finding the structure of the Virasoro module of local
martingales and comparing it with fusion products from CFT literature,
we attempt to interpret clearly in probabilistic terms the different SLE
variants used. Particularly the case of critical percolation
is well suited for both purposes, so we devote the whole
Section \ref{sec: kappa 6} to this.
On one hand SLE is known rigorously to describe the scaling limit
of the exploration path in percolation \cite{Smirnov-critical_percolation,
CN-critical_percolation_exploration_path}, while on the other hand
the question of the precise nature of the CFT of percolation
has attracted a lot of attention recently
\cite{MR-percolation_LCFT, RP-fusion_algebra_of_percolation,
RS-associative_algebraic_LCFT, EF-fusion_for_augmented_minimal_models}.
Despite definite successes such as the Cardy's crossing probability formula
\cite{Cardy-percolation},
fundamental questions about the conformal field theory appropriate for
percolation have remained unsettled,
notably that of the operator content of the theory.
In rough terms the operator content of a CFT should be a class of
representations (the fields of the theory)
closed under the fusion product. We will reproduce as
local martingales a few of the most fundamental fusions starting from
the Cardy's boundary changing field. Furthermore, we observe that the
simplest fusion that produces logarithmic representation appears in an SLE
variant that has a transparent probabilistic meaning: it is obtained by
conditioning on a crossing event. After such confirmations of the emerging
picture of the operator content of percolation \cite{MR-percolation_LCFT},
we speculate in Section \ref{sec: existence of Q21} whether
SLEs can be used to test another recent
argument: the claimed inconsistency of adding certain other operators to
the theory \cite{MR-logarithmic_M2p}.

The present exercise has a few byproducts perhaps
of direct interest to logarithmic conformal field theory. Most obviously,
contrary to the common situation in LCFT literature where logarithms appear
at specific rational values of the central charge, our logarithmic
representations for SLEs cover the entire
range $\infty < c \leq 1$. Section \ref{sec: ell 1} also shows
how extremely easy calculations give the values of logarithmic couplings,
and these results agree with and extend a particular case of
a recent proposal \cite{MR-logarithmic_M2p} made in the context of
logarithmic minimal models. It is also possible that our rather straighforward
setup may lend to development of alternative computational methods to
study fusion in CFTs.

\bigskip

We finish this introduction by briefly mentioning other works relating SLEs
and logarithmic conformal field theory.
While a coherent theory of such a relation is yet to be discovered,
the question has already been investigated with a few different
approaches. The first articles discussing SLE and LCFT in parallel
seem to be \cite{Rasmussen-SLE_and_LCFT,
MRR-logarithmic_conformal_null_vectors_and_SLE}.
More recently \cite{SPR-geometric_exponents} studied geometric questions
in lattice loop models that bear a relation to both
logarithmic minimal models and SLEs,
and \cite{CS-twist_operator_correlations}
used correlation functions of LCFT among other things to
provide a formula for the probability of an SLE event involving
a global geometric configuration of curves. A study of closed operator
algebras \cite{MR-logarithmic_M2p}
of LCFTs also exhibits a duality somewhat suggestive of SLEs.
Despite the fact that these few attempts start from
very different points of view, they all seem to point towards the conclusion
that the approapriate conformal field theory description of SLEs
is logarithmic. In hindsight we may observe that many of the
statistical models that originally required abandoning minimal models
in favor of logarithmic conformal field theory were of geometric nature
\cite{Saleur-conf_inv_polymers_and_percolation,
Cardy-logarithmic_correlations}:
polymers themselves or percolation clusters are objects whose global
spatial configuration is the natural object of interest.
In fact \cite{SPR-geometric_exponents} seems to suggest as some
sort of rule of thumb
that treating questions of global geometric kind, rather that just
correlations of local degrees of freedom at a distance, typically
requires logarithms in the CFT description
\cite{SPR-geometric_exponents}.

%  End Of Section  *************************************************
%  *****************************************************************

%  *****************************************************************
%  *  BACKGROUND  **************************************************
%  *****************************************************************
\section{Preliminaries}
\label{sec: background}

The aim is to make this paper readable for different audiences,
in particular knowledge of either SLE or CFT alone should suffice.
For this reason it is necessary to cover briefly a variety
of topics in this background section. Below are some suggestions about
accessible literature for each of the topics covered.
This section also serves to fix notations.

Section \ref{sec: SLE equations} is devoted to a brief account
on Schramm-Loewner evolutions (SLE), in particular variants known as
SLE${}_\kappa(\rho)$. The books and reviews \cite{Werner-random_planar_curves,
Lawler-conformally_invariant_processes, BB-2d_growth_processes,
KN-guide} serve as comprehensive treatments of the topic and contain
references to research articles.
The Virasoro algebra and the kinds of representations that will be
needed later on are discussed in Section \ref{sec: Virasoro representations}.
Classical results stated there can be found in e.g.
\cite{Astashkevich-Verma_modules},
whereas the more exotic staggered Virasoro modules are treated in detail
in \cite{KR-staggered}.
Finally, Section \ref{sec: local mgales} reminds in which sense
local martingales of the SLE growth process form a representation
of Virasoro algebra, a result taken from \cite{Kytola-local_mgales}.

\subsection{Schramm-Loewner evolutions in the upper half plane}
\label{sec: SLE equations}

The random curves described by the SLE growth processes are characterized
by two properties that make them suitable to describe curves in
statistical mechanics at criticality. One, the domain Markov property, is a
statement about the conditional probability law of the curve given an
initial segment of it, and it is often very easy to verify for curves
in statistical mechanics models. The other, conformal invariance, is a property
that is argued to emerge in the scaling limit taken at the critical
point.
Roughly speaking these properties imply that the local behavior of the
curve is characterized by one parameter, $\kappa > 0$, while precise
global statistics still depend on boundary conditions imposed.
Since %Riemann's mapping theorem states that
any two simply connected open domains (with at least two distinct
boundary points) of the Riemann sphere are conformally equivalent,
the conformal invariance property in particular allows us to work in a
reference domain, which we take to be
the upper half plane $\bH = \{ z \in \bC : \im z > 0 \}$.

\subsubsection{Chordal SLE${}_\kappa$ in the upper half plane}
\label{sec: chordal SLE def}

The simplest choice of boundary conditions is such that it merely ensures
the existence of a curve from one boundary point of the domain to another.
The corresponding SLE is known as the chordal SLE.
To define the chordal SLE${}_\kappa$ in $\bH$ from $0$ to $\infty$
consider the ordinary differential equation (Loewner's equation)
\begin{align}
\label{eq: Loewner}
\frac{\ud}{\ud t} g_t(z) \; = \; & \frac{2}{g_t(z)-X_t} \textrm{ ,}
\end{align}
with initial condition $g_0(z) = z \in \bH$.
The driving process $(X_t)_{t \in [0,\infty)}$ is taken to be
$X_t = \sqrt{\kappa} \; B_t$, where $(B_t)_{t \in [0,\infty)}$ is a standard
normalized Brownian motion and $\kappa > 0$.
For each $z \in \bH$, let $\tau_z$ be the
maximal time up to which the solution to (\ref{eq: Loewner}) exists
with the initial condition $z$.
The set $\bH_t = \{ z \in \bH : \tau_z > t\}$, $t \geq 0$, is
a simply connected open set and $g_t$ is the unique conformal map
$\bH_t \rightarrow \bH$ with $g_t(z) = z + \Order (z^{-1})$ as
$z \rightarrow \infty$. We denote the Laurent expansion at infinity
of $g_t$ by
\begin{align}
\label{eq: g expansion}
g_t(z) = z + \sum_{m=2}^\infty a_m(t) \, z^{1-m}
\textrm{ .}
\end{align}
It can be shown that in this setup the Loewner's equation
(\ref{eq: Loewner}) specifies a non self-crossing curve
$\gamma : [0,\infty) \rightarrow \overline{\bH}$
such that $\bH_t$ is the unbounded
component of $\bH \setminus \gamma[0,t]$, see \cite{RS-basic_properties}.
The curve $\gamma$ is called the trace of the chordal SLE${}_\kappa$
(in $\bH$ from $0$ to $\infty$).
For $\kappa \leq 4$ it is a simple curve (non-self intersecting),
for $\kappa > 4$ it has self-intersections, and for $\kappa \geq 8$ it
visits every point of the domain \cite{RS-basic_properties}.

For several lattice models of statistical physics,
with appropriate boundary conditions to account for a chordal curve,
the continuum limit of the law of a curve has been shown to be
chordal SLE${}_\kappa$ --- the value
of $\kappa$ depends on the particular model. For yet many others, similar
results are conjectured.
In Section \ref{sec: kappa 6} we will make the most explicit comparisons
with the CFT literature on percolation,
which corresponds to $\kappa=6$. The precise description of the continuum
limit result in this case will be given in
Section \ref{sec: exploration path}.

\subsubsection{SLE${}_\kappa(\rho)$ in the upper half plane}
\label{sec: SLE variants}

While for many lattice models it is possible to find boundary conditions such
that
the chordal SLE is the correct scaling limit, this is by no means the general
case. To account for somewhat more general boundary conditions, we take the
driving process $(X_t)$ to be a solution to the It\^o stochastic differential
equation
\begin{align}
\ud X_t \; = \; & \sqrt{\kappa} \; \ud B_t
    + \kappa \big( \pder{x} \log Z (X_t, g_t(Y_0)) \big) \; \ud t
\textrm{ ,}
\end{align}
where $(B_t)_{t \geq 0}$ is standard Brownian motion
and $Z(x,y) = (x-y)^{\rho/\kappa}$ is what we call the partition function.
The processes are defined on a random
time interval $t \in [0,\tau)$, where $\tau$ is a stopping time.
The resulting process is known as SLE${}_\kappa (\rho)$, and it is a
variant of SLE in the sense that the probability measure is absolutely
continuous with respect to the chordal SLE${}_\kappa$ when restricted to
the curve up to times smaller than $\tau$.
One could allow more general variants by letting the partition function
$Z$ depend on more points, e.g. \cite{Kytola-local_mgales}. Although
mostly SLE${}_\kappa(\rho)$ is enough for us, we will once in Section
\ref{sec: existence of Q21} replace the $Z$ above by
$Z(x,y_1,y_2) = (x-y_1)^{\rho_1/\kappa} (x-y_2)^{\rho_2/\kappa}
(y_1-y_2)^{\rho_1 \rho_2/2\kappa}$, and get a variant which we call
SLE${}_\kappa(\rho_1, \rho_2)$. In the rest of the article we denote briefly
$g_t(Y_0) =: Y_t$.

\subsubsection{Examples of SLE${}_\kappa(\rho)$}
\label{sec: SLE variant examples}

Without even considering lattice models, it is easy to see that the
generalization SLE${}_\kappa(\rho)$ is useful. As an obvious remark we note
that with $\rho=0$ the partition function is constant $Z(x,y)=1$, thus
$\pder{x} \log Z = 0$, and SLE${}_\kappa(\rho=0)$ is nothing but the
chordal SLE.\footnote{The
statement of equality of chordal SLE${}_\kappa$ and SLE${}_\kappa(\rho=0)$
means that the statistics of the curves $\gamma$ are identical. More precisely,
since SLE${}_\kappa(\rho)$ need not be defined for all $t \geq 0$,
the correct interpretation is that the law of the curves should be considered
up until stopping times up to which both are defined.
There is another minor difference in the definitions. In
SLE${}_\kappa(\rho)$ we have another marked point $Y_0$
at which a boundary condition changing field is present. Even at $\rho=0$
when this has no effect on statistics of $\gamma$, we choose to keep track
of $Y_t=g_t(Y_0)$.}
Below we will give two more interesting examples of SLE${}_\kappa(\rho)$,
which will also facilitate giving probabilistic interpretations in later
sections.

First, take the chordal SLE${}_\kappa$, for $\kappa>4$,
and imagine conditioning it on the event that the curve doesn't touch the
interval $[Y_0,\infty)$, a situation that has been considered e.g. in
\cite{Dubedat-duality}.
Equivalently, the difference of the $X$ and $Y$ processes,
$D_t = Y_t-X_t$, shouldn't hit zero.
The process $D_t$ is a linear time change of Bessell process of dimension
$d=\frac{4+\kappa}{\kappa}$.
One easily computes by It\^o's formula that starting from $D_0$, the
probability that $|D_t|>0$ for all $t \in [0,T]$ is
$\PR \big[ |D_t|>0 \textrm{ for } 0 \leq t \leq T \big] \, = \,
G(|D_0|/\sqrt{T})$,
where
\begin{align*}
G(r) \; = \; 
\frac{\int_0^r s^{1-d} e^{-\frac{1}{2\kappa} s^2} \; \ud s}
{\int_0^\infty s^{1-d} e^{-\frac{1}{2\kappa} s^2} \; \ud s}
\textrm{ .}
\end{align*}
Under the conditioned probability measure
$\PR \big[ \; \cdot \; \big| |D_t| > 0 \textrm{ for } t \in [0,T] \big]$,
the driving process $X_t$ satisfies by Girsanov's formula
\begin{align*}
\ud X_t \; = \; \sqrt{\kappa} \; \ud B'_t
    + \kappa \; \Big( \pder{D} \; \log G \big( \frac{|D_t|}{\sqrt{T}} \big)
    \Big) \; \ud t
\end{align*}
where $B'$ is a standard Brownian motion under the conditioned measure.
Taking $T \rightarrow \infty$, this tends to
\begin{align*}
\ud X_t \; = \; \sqrt{\kappa} \; \ud B'_t
    + \frac{\kappa-4}{X_t-Y_t} \; \ud t
\end{align*}
which is the driving process of SLE${}_\kappa(\rho)$ with
$\rho = \kappa-4$. Thus indeed a simple conditioning of the chordal
SLE${}_\kappa$ leads to an SLE${}_\kappa(\rho)$.

As the second example, we will consider coordinate changes. Suppose
we take an SLE${}_\kappa(\rho')$ in $\bH$
(in particular $\rho'=0$ corresponds to the chordal SLE) and consider a
M\"obius image of the curve. More precisely, take $\mu : \bH \rightarrow \bH$
a M\"obius transformation of the half plane such that
$\mu(X_0) = \tilde{X}_0 \in \bR$, $\mu(Y_0) = \infty$ and
$\mu(\infty) = \tilde{Y}_0 \in \bR$. Then consider the curve
$\tilde{\gamma} = \mu(\gamma)$ as an unparametrized but oriented curve
starting from $\tilde{X}_0$. A standard computation (see e.g.
\cite{SW-coordinate_changes}) shows that the law of $\tilde{\gamma}$ is
SLE${}_\kappa(\rho=\kappa-6-\rho')$.
For example the chordal SLE${}_\kappa$ in $\bH$ whose target point is
$Y_0$ instead of $\infty$ is obtained as the above
M\"obius image (because of the conformal invariance property),
therefore it is an SLE${}_\kappa(\rho=\kappa-6)$.
We could also coordinate change the first example $\rho'=\kappa-4$.
A chordal SLE${}_\kappa$ from $X_0$ to $Y_0$ conditioned not to touch
$(Y_0,\infty]$ is then seen to be an SLE${}_\kappa(\rho=-2)$.
Interestingly for the purpose of the present
paper, this last example will always exhibit logarithms as will be shown
in Section \ref{sec: ell 0}.

\subsection{Virasoro algebra and its representations}
\label{sec: Virasoro representations}
The Virasoro algebra $\vir$ is the complex Lie algebra spanned by
$L_n$, $n\in \bZ$ and $C$ with commutation relations
\begin{align*}
[L_n , L_m] \; = \;
    (n-m) \; L_{n+m} + \frac{n^3-n}{12} \delta_{n+m,0} \; C
\quad , \qquad [C, L_n] \; = \; 0
\textrm{ .}
\end{align*}
In a conformal field theory the central charge $c$ is a number
characteristic of the theory such that
$C$ acts as $c \, \unit$ in all the representations. We will
only consider such representations and the value of $c$ is determined
by $\kappa$ via $c(\kappa) = 13 - 6 \, \frac{\kappa}{4}
- 6 \, \frac{4}{\kappa}$.
Let $\sU$ denote the universal enveloping algebra of $\vir$ modulo the
identification of $C=c\unit$ and
$\sU^\pm$ the universal enveloping algebras of the Lie subalgebras
$\vir^\pm$ generated by $L_n$ with $\pm n > 0$. These are graded
by $\sU = \oplus_{m \in \bZ} \; \sU_m$, where the homogeneous component
$\sU_m$ is has a Poincar\'e-Birkhoff-Witt basis consisting of
$\cdots L_{-2}^{k_{-2}} \, L_{-1}^{k_{-1}} \, L_{0}^{k_0}
\, L_{1}^{k_1} \, L_{2}^{k_2} \cdots$ with $\sum_{n \in \bZ} n k_n = - m$,
and similarly for $\sU^\pm$.

We fix some notation and discuss the class of representations we will
consider. For a $\vir$ module $\sM$, let $\sM_{[\eta]}$ be the subspace
consisting of those $u \in \sM$ for which $(L_0 - \eta)^N u = 0$ for
$N$ large enough. We call $\sM_{[\eta]}$ the generalized eigenspace
of eigenvalue $\eta$.

The representations we will study can be written as
direct sums of finite dimensional generalized $L_0$
eigenspaces, 
with eigenvalues bounded from below. In indecomposable
representations the generalized eigenvalues can only differ by
integers, so denoting by $h$ the lowest $L_0$ eigenvalue and
by $\sM_m = \sM_{[h+m]}$ we have
$\sM = \oplus_{m=0}^\infty \, \sM_m$. We call $\sM_m$ the grade $m$ of $\sM$.
Note that $L_n$, or indeed any $U \in \uea_{-n}$, maps $\sM_{m}$ to $\sM_{m-n}$.
To any such module $\sM$ we can associate the contragredient module $\sM^*$
as follows. Let $\sM_n^*$ be the vector space dual of the finite
dimensional space $\sM_n$ and define $\sM^* = \oplus_{n=0}^\infty \sM_n^*$,
with Virasoro action given by
$\bra L_n \, v^*, v \ket = \bra v^* , L_{-n} \, v \ket$.
To facilitate a similar formula with the universal enveloping algebra
the anti-involution $\dagger$ is defined by linear extension of
$(L_{n_1} \cdots L_{n_k})^\dagger \, = \, L_{-n_k} \cdots L_{-n_1}$,
so that $\bra U^\dagger \, v^*, v \ket = \bra v^* , U \, v \ket$ for all
$U \in \sU$, $v \in \sM$, $v^* \in \sM^*$.

\subsubsection{Highest weight representations}
The most important representations for conformal field theory are
highest weight representations. A $\vir$ module $\sH$ is called
a highest weight representation of highest weight $h \in \bC$ if there
exists a non-zero $\ghwv \in \sH$, called a highest weight vector, such
that $L_0 \, \ghwv = h \, \ghwv$, %$C v_0 = c v_0$
$L_n \, \ghwv = 0$ for all $n>0$ and $\sH = \sU \, \ghwv$.
There is a universally repelling object in the category of highest
weight modules of fixed $c$ and $h$, the Verma module with highest
weight vector $v_{c,h}$
\begin{align*}
\Verma_{c,h} \; = \; \bigoplus_{\substack{k \in \bN \\
        0 < n_1 \leq n_2 \leq \cdots \leq n_k}}
    \bC \; L_{-n_k} \cdots L_{-n_1} \; v_{c,h}
\textrm{ .}
\end{align*}
The universal property guarantees that for any
highest weight module $\sH$ with a highest weight vector $\ghwv$
there exists a unique surjective $\vir$-homomorphism
$\Verma_{c,h} \rightarrow \sH$ such that $v_{c,h} \mapsto \ghwv$.
Therefore any highest weight module is a quotient of the Verma module
by a submodule.

The submodule structure of Verma modules was resolved by Feigin and
Fuchs \cite{FF-Verma_modules_1983, FF-Verma_modules_1984}, see also
\cite{FF-representations, Astashkevich-Verma_modules}.
We will state the result below, emphasizing the differences in the
cases when $\kappa$ is rational or not.
Recall first that $c$ is parametrized as
$c=c(\kappa)=13 - 6 \frac{\kappa}{4} - 6 \frac{4}{\kappa}$.
In addition we define
$h_{r,s}(\kappa) = \frac{\kappa/4}{4} (r^2-1) - \frac{1}{2} (rs-1) +
\frac{4/\kappa}{4} (s^2-1)$.
We point out that this parametrization has the symmetries
$h_{-r,-s} = h_{r,s}$ and $h_{r+a,s+a\kappa/4} = h_{r,s}$ for any $a$,
and also note that $h_{r,s}+rs = h_{r,-s}$.

\begin{figure}
\begin{center}
\includegraphics[width=1.0\textwidth]{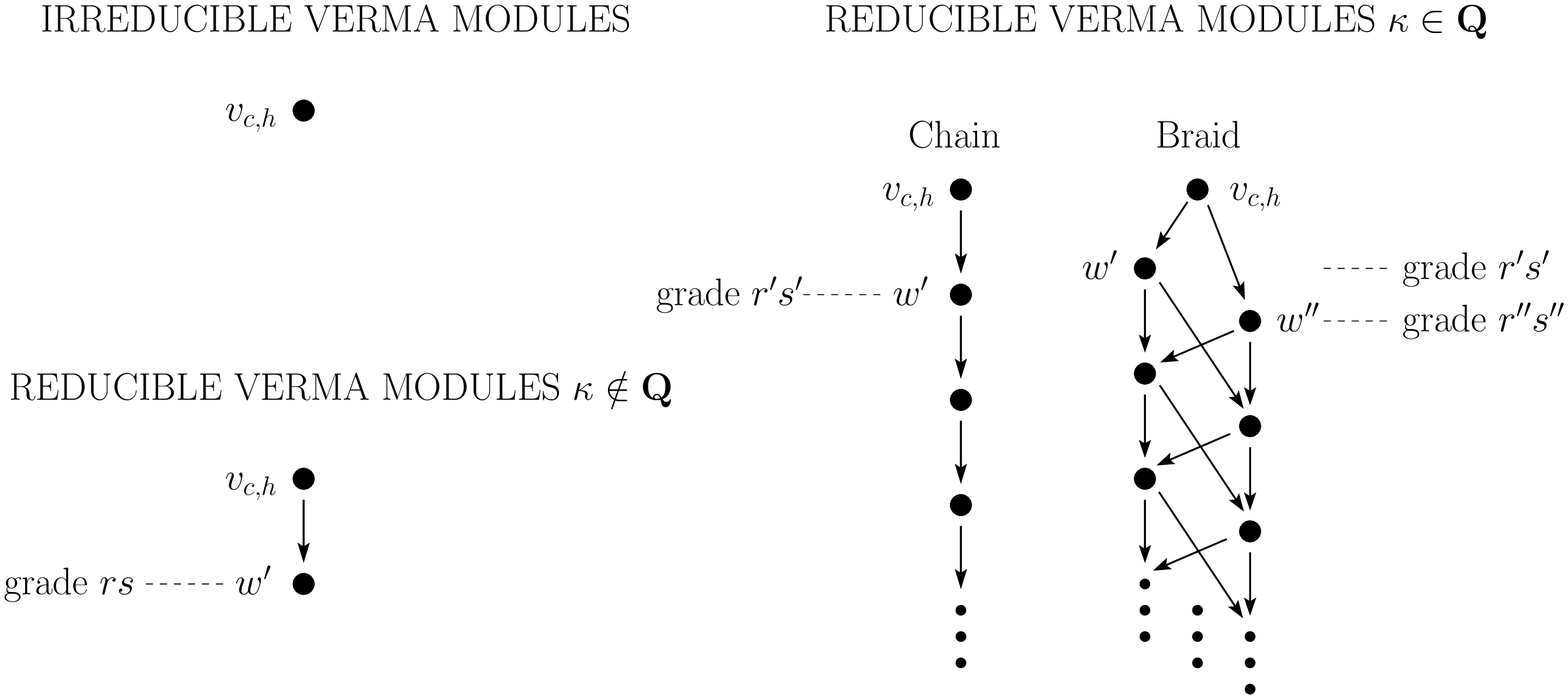}
\caption{
\emph{Submodule structure of Verma modules.}}
\label{fig: Verma module}
\end{center}
\end{figure}

The most important thing about the submodules of the Verma module
$\Verma_{c(\kappa),h}$ is that any submodule is generated by singular vectors,
eigenvectors $\ghwv'$ of $L_0$ which are annihilated by the positive
generators, $L_n w' = 0$ for $n>0$
(such a vector generates a highest weight submodule). Furthermore,
up to multiplicative scalar there is never more than one singular vector at a
given grade.
It can be shown that there are no other singular vectors than those
proportional to the highest weight vector $v_{c,h}$ in the
Verma module $\Verma_{c(\kappa),h}$
unless $h = h_{r,s}(\kappa)$ for some $r,s \in \bZ_+$.
In other words, $\Verma_{c(\kappa),h}$ can only be reducible if
$h=h_{r,s}(\kappa)$. We will describe these reducible cases below and in
and the Figure \ref{fig: Verma module}, where the dots represent singular
vectors and an arrow from one dot to another signifies the fact that the
latter is contained in the submodule generated by the former.

\bigskip

\emph{Irrational $\kappa$:} Suppose $\kappa \notin \bQ$ and
$h = h_{r,s}(\kappa)$ for some $r,s \in \bZ_+$. The choice of such $r,s$
is unique when $\kappa$ is not rational. Then there is a singular vector
$\ghwv'$
in $\Verma_{c(\kappa),h}$ at grade $rs$. The submodule generated by $\ghwv'$
is the maximal proper submodule of $\Verma_{c(\kappa),h}$ and is itself
an irreducible Verma module, $\uea \ghwv' \isom \Verma_{c(\kappa), h+rs}$.

\smallskip

\emph{Rational $\kappa$:} Suppose $\kappa = 4 \frac{p}{q}$ with
$p,q \in \bZ_+$ relatively prime. Suppose furthermore that
$h=h_{r,s}(\kappa)$ for some $r,s \in \bZ_+$. It is convenient to
to take $r',s'$ such that $h=h_{r',s'}(\kappa)=h_{r,s}(\kappa)$ and
the product $r's'>0$ is minimal.
There are still two structural possibilities
that we name "braid" and "chain" following the terminology introduced
in \cite{KR-staggered}.

Assume that $q \nmid r$ and $p \nmid s$, a case that we will call
\emph{braid}. Then we choose also $r'',s'' \in \bZ_+$ such that
$h=h_{r,s}(\kappa)=h_{r'',s''}(\kappa)$ and $r'' s'' > r' s'$ is minimal
among the remaining choices.
The Verma module $\Verma_{c(\kappa),h}$ contains singular
vectors $\ghwv'$ at grade $r' s'$ and $\ghwv''$ at grade $r'' s''$ and the maximal
proper
submodule is generated by these. The highest weight submodules generated by
these singular vectors are themselves braid type Verma modules
$\uea \ghwv' \isom \Verma_{c(\kappa), h+r's'}$ and
$\uea \ghwv'' \isom \Verma_{c(\kappa),h+r''s''}$. The intersection
$\uea \ghwv' \cap \uea \ghwv''$ is the maximal proper submodule in both
$\uea \ghwv'$and $\uea \ghwv''$.

The other possibility, $q \mid r$ or $p \mid s$, will be called \emph{chain}.
There is a singular vector $\ghwv'$ at grade $r's'$ and $\uea \ghwv'$ is the
maximal
proper submodule of $\Verma_{c(\kappa),h}$. This submodule is itself a chain
type Verma module, $\uea \ghwv' \isom \Verma_{c(\kappa), h+r's'}$.

\bigskip

We adopt a few notational conventions for highest weight modules.
Verma modules $\Verma_{c,h}$ were already considered above.
Other highest weight modules are quotients of the Verma modules. In pictures
we use the convention that a black (filled) dot represents a non-zero
singular vector whereas a white (empty) dot represents a singular vector
that is quotiented away from the Verma module (we sometimes refer to this as a
null vector).
Arrows still represent inclusions between the submodules generated by the
singular vectors.
We denote by $\sQ_{\kappa ; r,s}$ the
highest weight module that is the quotient of
$\Verma_{c(\kappa), h_{r,s}(\kappa)}$ by the
submodule generated by singular vector at level\footnote{If $r,s \in \bZ_+$
there always exists a singular vector at grade $rs$, even if $rs>0$ is not
minimal.} $rs$, and by
$\ghwv_{\kappa;r,s}$ %$ = v_{c,h_{r,s}} + \sV_{c,h_{r,s}+rs}$
the highest weight vector in $\sQ_{\kappa; r,s}$.
In general $\sQ_{\kappa;r,s}$ need not be irreducible (it is only if
$\kappa \notin \bQ$ or if in the chain case $rs>0$ is minimal).
%For $\kappa \notin \bQ$ this is irreducible
We denote the irreducible highest weight module obtained
as the quotient of $\Verma_{c, h}$ by its maximal proper
submodule by $\sL_{c,h}$.
For notational simplicity we also often omit explicit references to $\kappa$
and $c$.

\subsubsection{Staggered modules}
\label{sec: staggered theory}

In logarithmic conformal field theories one encounters indecomposable
representations more complicated than just highest weight representations.
In particular, the property that is responsible for the logarithms
in correlation functions is the
non-diagonalizability of $L_0$ \cite{Gurarie-logarithmic_operators}.
For our purposes, staggered modules
\cite{Rohsiepe-reducible_but_indecomposable, KR-staggered} of the
kind that we describe below will be enough. We remark that in Sections
\ref{sec: representations} -- \ref{sec: kappa 6} we will study
concretely constructed $\vir$-modules, so it is not
important for us to know beforehand when a module of certain structure
exists. Instead it is important to know which information is sufficient
to uniquely characterize the module.
The results stated below can be found in
\cite{KR-staggered}, see also \cite{Rohsiepe-reducible_but_indecomposable}.

Suppose $\sH^\lft,\sH^\rgt$ are two highest weight modules of the
same central charge $c$ and highest weights $h^\lft, h^\rgt$ and
highest weight vectors $w^\lft, w^\rgt$ respectively.
We call $\Stagg$ a staggered module with left module $\sH^\lft$ and
right module $\sH^\rgt$ if it contains the former as a submodule
$\sH^\lft \subset \Stagg$, its quotient by this
submodule is the latter $\Stagg / \sH^\lft \isom \sH^\rgt$,
and $L_0$ is not diagonalizable on $\Stagg$.
This may be summarized in the short exact sequence
$$0 \longrightarrow \sH^\lft \overset{\iota}{\longrightarrow} \Stagg
\overset{\pi}{\longrightarrow} \sH^\rgt \longrightarrow 0$$
in addition to which we have the requirement of non-diagonalizability
of $L_0$.

The simplest case is when $h^\rgt = h^\lft$.
Then we set $\xi = \iota(w^\lft)$ and pick a representative
$\eta \in \pi^{-1} (\bC w^\rgt)$ such that $(L_0 - h^\rgt) \eta = \xi$
(possible by non-diagonalizability of $L_0$). The structure of such a
staggered module is determined by the mere knowledge of $\sH^\lft$
and $\sH^\rgt$.\footnote{However the other direction, existence of such a
staggered module for given $\sH^\lft$ and $\sH^\rgt$ is more delicate ---
see \cite{KR-staggered} for conditions characterizing the existence.}

A more interesting case is if $h^\rgt = h^\lft + \ell$ with $\ell \in \bZ_+$.
We can then choose $\xi = \iota(w^\lft)$ and
$0 \neq \eta \in \pi^{-1}(\bC w^\rgt)$ in the generalized $L_0$ eigenspace of
eigenvalue $h^\rgt$. Denote $(L_0-h^\rgt) \eta = \omega_0 \in \iota(\sH^\lft)$
and assume without
loss of generality $\omega_0 \neq 0$ (were this not possible, $L_0$ would have
to be diagonalizable). For $n>0$ we have
$L_n \omega_0 = L_n (L_0 - h^\rgt) \eta = (L_0 - h^\rgt + \eta) L_n \eta = 0$,
since $L_n \eta \in \sH^\lft_{\ell-n}$.
What this means is that $\omega_0$ is a non-zero singular vector in
$\iota(\sH^\lft)$,
and in particular $\ell$ must be a grade of a singular vector in the Verma
module of highest weight $h^\lft$. We can assume
a choice of $\eta$ such that $\omega_0 = \chi \xi$ where
%$\chi \in \uea^-_{\ell}$ is of the form
\begin{align*}
\chi = L_{-1}^{\ell} +
\sum_{\substack{n_1 \geq n_2 \geq \cdots \geq n_k \geq 1 \\
        n_1 > 1 \; , \; \sum_j n_j = \ell}}
d_{n_1, \ldots, n_k} \; L_{-n_1} \cdots L_{-n_k}
\; \in \; \sU^-_{\ell}
\textrm{ .}
\end{align*}
Unlike the case $h^\rgt=h^\lft$, here the structure of the module is in
general not yet determined by the knowledge of left and right modules
$\sH^\lft, \sH^\rgt$. It can be shown that depending on the particular
$\sH^\lft, \sH^\rgt$ the set of isomorphism classes of such staggered modules
is an affine subspace (possibly empty) of a vector space of dimension
at most two \cite{KR-staggered}.
Below we will content ourselves to state what is enough to identify
the modules in a case that suffices for the needs of this paper.

Suppose the highest weight of $\sH^\lft$ is $h^\lft=h_{r,s}$ with
$r,s \in \bZ_+$ such that the grade $\ell = rs > 0$ the minimal at which
$\sH^\lft$ contains a nonzero singular vector.
With the notation and normalizations as above, note that
$\chi^\dagger \eta$ must be proportional to the highest weight vector of
$\iota(\sH^\lft)$, that is $\chi^\dagger \eta = \beta \xi$ for some
$\beta \in \bC$. This $\beta$ is invariant (independent of the
choices made), and given $\sH^\lft, \sH^\rgt$ its value is enough to determine
the structure of the staggered module $\Stagg$.
The value of $\beta$ is known also as \emph{logarithmic coupling}.

\subsection{The Virasoro module of SLE local martingales}
\label{sec: local mgales}

We now briefly recall the result of \cite{Kytola-local_mgales} by
which local martingales of the SLE growth process form a Virasoro
module.

Let $c, h_x, h_y \in \bC$ be parameters. Consider independent
formal variables $a_2, a_3, \ldots$.
The differential operators %$L_n$, $n \in \bZ$,
on $\sF = C^\infty(\{(x,y) \in \bR^2 : x \neq y\}) [a_2, a_3, a_4, \ldots]$
given below satisfy the commutation relations of the Virasoro algebra
\begin{align*}
L_2 \; = \; & -\pder{a_2} + \sum_{l \geq 4} \big( (l-3) a_{l-n} \big)
    \pder{a_{l}} \\
L_1 \; = \; & \sum_{l \geq 3} \big( (l-2) a_{l-1} \big)
    \pder{a_l} + \pder{x} + \pder{y} \\
L_{n} \; = \; & \frac{1}{(n-2)!}
    \Big[ \big[ [ \cdot \! \cdot \! \cdot [L_2,
    \underbrace{L_1], \ldots ], L_1 \big], L_1}_{\textrm{$n-2$ times}} \Big]
\quad \textrm{ for $n>2$}
\end{align*}
\begin{align*}
L_0 \; = \; & \sum_{l \geq 2} l a_{l} \pder{a_l}
    + x \pder{x} + y \pder{y} + h_{x} + h_{y}
\end{align*}
\begin{align*}
L_{-1} \; = \; & \sum_{l \geq 2} \Big( (l+2) a_{l+1} +
    \sum_{\substack{ m_1, m_2 \geq 2 \\ m_1+m_2 = l+1}} a_{m_1} a_{m_2}
    \Big) \pder{a_l} \\
& \qquad + \big( x^2 - 3 a_{2} \big) \pder{x} 
    + \big( y^2 - 3 a_{2} \big) \pder{y} + 2 x h_{x} + 2 y h_{y} \\
L_{-2} \; = \; & \sum_{l \geq 2} \Big( (l+4) a_{l+2} - 4 a_{2} a_{l}
    + \sum_{\substack{ m_1, m_2 \geq 2 \\ m_1+m_2 = l+2}} 3 a_{m_1} a_{m_2} \\
& \qquad \qquad \qquad
    + \sum_{\substack{ m_1, m_2, m_3 \geq 2 \\ m_1+m_2+m_3 = l+2}}
    a_{m_1} a_{m_2} a_{m_3} \Big) \pder{a_l} \\
& \qquad + \big( x^3 - 4 x a_{2} -5 a_{3} \big) \pder{x}
    + \big( y^3 - 4 y a_{2} -5 a_{3} \big) \pder{y} \\
& \qquad + (3 x^2 - 4 a_{2}) h_{x} + (3 y^2 - 4 a_{2}) h_{y}
    - \frac{c}{2} a_{2} \\
L_{-n} \; = \; & \frac{1}{(n-2)!} \Big[
    \underbrace{L_{-1}, \big[ L_{-1}, [ \ldots , [L_{-1}}_{\textrm{$n-2$ times}}
              ,L_{-2}] \cdot \! \cdot \cdot ] \big] \Big]
\quad \textrm{ for $n>2$.}
\end{align*}
We remark that as Virasoro modules are graded by their $L_0$ eigenvalues,
it is reasonable in view of the explicit expression above to consider
the variables $x$ and $y$ having degree $1$ and the variable $a_l$ having
degree $l\geq2$. The homogeneity degree of a function $\varphi \in \sF$
then simply differs from its $L_0$ eigenvalue by %the additive constant
$h_x+h_y$.

Let $\varphi \in \sF$
and consider the stochastic process
$\varphi_t = \varphi(X_t, Y_t, a_2(t), a_3(t), \ldots)$ where
$X_t, Y_t, a_2(t), a_3(t), \ldots$ are the processes for SLE${}_\kappa(\rho)$
as defined in Section \ref{sec: SLE equations}.
Then It\^o's formula says that
$$Z(X_t, Y_t) \; \ud \big( \frac{\varphi_t}{Z(X_t,Y_t)} \big)
= (\cdots) \; \ud B_t + (A_{\kappa; \rho} \; \varphi) (X_t, Y_t, a_2(t), \ldots)
\; \ud t \textrm{ ,}$$
where $A_{\kappa; \rho}$ is the differential operator
\begin{align*}
A_{\kappa; \rho} \; = \; & \frac{\kappa}{2} \; \frac{\partial^2}{\partial x^2} +
    \frac{2}{y - x} \; \pder{y}
    - \frac{\rho(\rho+4-\kappa)/ (2\kappa)}{(y-x)^2} \\
& \qquad 
+ 2 \sum_{l \geq 2} \Big( \sum_{\substack{0 \leq m \leq l-2 \\
                0 \leq r \leq \lfloor \frac{l-2-m}{2}\rfloor}} x^m 
    (-1)^r \frac{(m+r)!}{m! \; r!}
    \sum_{\substack{k_1, \ldots k_r \geq 2 \\ \sum k_j = l-m-2}}
    a_{k_1} \cdots a_{k_r} \Big) \; \pder{a_l}
\end{align*}
and $Z(x,y) = (x-y)^{\rho/\kappa}$. We call
$\Kern A_{\kappa; \rho} \subset
C^\infty(\{(x,y) : x \neq y\}) [a_2, a_3, a_4, \ldots]$
the space of local martingales of SLE${}_\kappa(\rho)$,
because for any $\varphi \in \Kern A_{\kappa; \rho}$,
the process $\frac{\varphi_t}{Z(X_t, Y_t)}$ is a local martingale.

If $c = c(\kappa) = 13 - 6 \frac{\kappa}{4} - 6\frac{4}{\kappa}$,
$h_x = h_{1,2}(\kappa) = \frac{6-\kappa}{2\kappa}$ and
$h_y = \frac{\rho(\rho+4-\kappa)}{4\kappa}$, then
$[L_n,A_{\kappa; \rho}] = q_n(x,a_2,a_3,\ldots) \; A_{\kappa; \rho}$,
where $q_n$ is a polynomial
multiplication operator. In particular the space of local martingales,
$\Kern A_{\kappa; \rho}$, is a Virasoro module.
From now on assume that $c, h_x, h_y$ are fixed in this way.

\bigskip

The physical interpretation of the above representation is
the action of local conformal transformations
on the boundary changing field located at infinity.
The key idea in finding the representation was the idea of an
``SLE state'' built by composing intertwining operators corresponding to
all the boundary changes on the real axis, followed by an implementation of
the conformal transformation \cite{BB-conformal_transformations,
Kytola-local_mgales}.
One may therefore expect the representation to be closely related to the
fusion product of the boundary condition changing fields, which is indeed
what we will see in the examples.

\bigskip

In the rest of the paper we will study the structure
of this Virasoro module of local martingales for different SLE variants.
In all cases, the modules $\Kern A_{\kappa;\rho}$ turn out to be
essentially contragredient to the fusion products of the boundary changing
fields.
In addition to giving the results about the module $\Kern A_{\kappa;\rho}$,
we always attempt both to discuss what
is known about the corresponding
fusion, and to give a probabilistic interpretation of the SLE variant used.

Once the question is posed this way and the SLE variant in question fixed,
the study of the representations has become a very concretely posed problem.
Keeping in mind generalities from representation theory,
one typically needs to answer explicit questions such as computing graded
dimensions (characters), checking whether certain descendants
of vectors vanish or belong to subrepresentations etc. Of course
even these checks may soon become laborous.
We have performed them with the help of a computer by implementing
computations with Virasoro algebra and the explicit
differential operators $L_n$ and $A_{\kappa; \rho}$ given above.

%  End Of Section  *************************************************
%  *****************************************************************

\section{Example modules and appearance of logarithms}
\label{sec: representations}

\subsection{The module for chordal SLE${}_\kappa$}
\label{sec: chordal SLE representation}

Before starting to analyze the more complicated cases,
it is worth taking a look at the simplest and already well understood case,
the chordal SLE${}_\kappa$.
This doesn't yet involve fusions because the only boundary
changing field on real axis is at the starting point of the curve.
However even this simple case is quite interesting:
for example in the case $\kappa=6$ discussed in Section \ref{sec: kappa 6},
we will find a representation
contragredient to the conformal family of the field $\psi$ Cardy used
in deriving his famous crossing probability formula.

In the case of chordal SLE${}_\kappa$,
the partition function should be thought of as a constant $Z = 1$.
The space of local martingales, that is the kernel of
\begin{align*}
A_{\kappa} \; = \; \frac{\kappa}{2} \; \frac{\partial^2}{\partial x^2}
%    + \sum_{m \geq 2} \Res{w} \frac{2 w^{m-2}}{f(w)-x}
%    \; \pder{a_m}
+ 2 \sum_{l \geq 2} \Big( \sum_{\substack{0 \leq m \leq l-2 \\
                0 \leq r \leq \lfloor \frac{l-2-m}{2}\rfloor}}
    \frac{x^m (-1)^r (m+r)!}{m! \; r!}
    \sum_{\substack{k_1, \ldots k_r \geq 2 \\ \sum k_j = l-m-2}}
    a_{k_1} \cdots a_{k_r} \Big) \; \pder{a_l}
\textrm{ ,}
\end{align*}
has been studied in \cite{BB-SLE_martingales}. The differential
operator $A_\kappa$ preserves the space
$\sF^\poly = \bC[x,a_2,a_3,a_4,\ldots]$
of polynomials. A clever argument allowed Bauer and Bernard to show
that $A_\kappa$ maps $\sF^\poly_m$ surjectively to $\sF^\poly_{m-2}$,
where the homogeneous subspaces are defined as before by $L_0$ eigenvalues.
Combined with the observation
\begin{align*}
\dmn \sF^\poly_{m} \; = p(m) \; := \; &
\# \{ (k_1, \ldots, k_m) \in \bN^m \; : \; \sum_j j k_j = m\} %\\
%= \; & \# \{ 0 < n_1 \leq \cdots \leq n_k \; : \;
%    k \in \bN \; , \; \sum_j n_j = m \}
\textrm{ ,}
\end{align*}
the surjectivity
shows that
the graded dimension of $\Kern A_\kappa \subset \sF^\poly$ coincides with
that of $\sQ_{\kappa;1,2}$, namely $\dmn (\Kern A_\kappa)_{m} = p(m) - p(m-2)$.
Better yet, a concrete construction \cite{BB-conformal_transformations}
shows for any chordal SLE${}_\kappa$ that $\Kern A_\kappa$ is contragredient
to the highest weight representation $\sQ_{\kappa ; 1,2}$. The
vectors of the lowest $L_0$ eigenvalue $h_{1,2}(\kappa)$ are constants,
corresponding to the partition function of chordal SLE, $Z(x)=1$.
This last remark relates to the trivial observation that
$Z \in \Kern A_{\kappa;\rho}$, or the fact that constant processes are
(local) martingales.

\bigskip

We will start by discussing the case of irrational $\kappa$
%in Sections \ref{sec: ell 0}, \ref{sec: ell 1} and \ref{sec: ell 2},
as it is the simplest in terms of representation theory.
In this case $\sQ_{\kappa;1,2}$ is the irreducible highest
weight representation
of highest weight $h = h_{1,2}(\kappa) = \frac{6 - \kappa}{2 \kappa}$.
Now $\Kern A_\kappa \subset \sF^\poly$ is
also the irreducible $\sQ_{\kappa;1,2} = \sL_{h_{1,2}}$
with the partition function $Z = 1$ as its highest weight vector,
as can be seen directly from the graded dimension as in
\cite{BB-SLE_martingales} or as follows from the fact that an irreducible
highest weight representation is contragredient to itself.
We emphasize that the situation will be different in the case of rational
$\kappa$: when $\sQ_{1,2}$ is a reducible highest weight representation,
the contragredient $\sQ_{1,2}^*$ is no longer a highest weight representation.
The particular case of $\kappa=6$ will be made explicit in Section
\ref{sec: Cardys field}.

\subsection{Naive fusions and the appearance of logarithms}
\label{sec: naive fusion}

The purpose of this paper is to exhibit logarithmic representations in the
context of SLEs. The characteristic property of these representations is a
nontrivial Jordan block structure
of $L_0$. This is a degeneracy that can result from coincidence of
conformal weights of several primary fields. This section is a discussion
of how such coincidences appear in fusions.

While a proper definition of fusion product of representations is quite
delicate \cite{Gaberdiel-fusion_in_CFT}, there is also a somewhat
naive, but very straightforward way of understanding fusion in CFT.
This consists of saying that a primary field of conformal weight $h$
occurs in the fusion product of
primary fields of weights $h'$ and $h''$ if the three point function of
these three fields is non-vanishing.\footnote{The idea is that fusion is
the short distance expansion (OPE) of two fields, and the presence in this
expansion of any other field whose two point function is nonzero, can be
tested by taking the correlation function of the short distance expansion
with this other field.} Since
three point functions are up to a multiplicative constant fixed by
conformal invariance,
the question boils down seeing whether a nonzero three point function
satisfies all other
requirements --- notably the null field equations (differential equations
that one gets via Ward identities \cite{DMS-CFT}) if either of the fused
fields has a vanishing descendant.

For concreteness, motivated by the physical understanding of
the boundary changing fields for SLE let us consider naively the fusion of
a field of weight
$h_{1,2} = \frac{6-\kappa}{2 \kappa}$ with a vanishing descendant at level
two, and a field of weight
$h(\rho) = \frac{1}{4 \kappa} \rho (\rho + 4 - \kappa)$. The three
point function of these two located at $x$ and $y$ with a field of weight
$h$ at infinity
would be proportional to $(x-y)^{h-h_{1,2}-h(\rho)}$. The appropriate null
field equation relates immediately to SLE local martingales: it in fact reads
$A_{\kappa; \rho} \, \big( (x-y)^{h-h_{1,2}-h(\rho)} \big) = 0$.
This boils down to the following equation for the conformal weights
\begin{align} \label{eq: naive fusion}
\big( h - h_{1,2} - h(\rho) \big) \big( h - h_{1,2} - h(\rho) - 1 \big)
    + \frac{4}{\kappa} \big( h - h_{1,2} - h(\rho) \big)
    - \frac{4}{\kappa} h(\rho) \; = \; 0
\textrm{ .}
\end{align}
The two roots correspond naively to the two conformal weights that
appear in the fusion
\begin{align*}
h^+ \; = \; & \frac{\rho^2 + 8 \rho - \rho \kappa + 12 - 2 \kappa}{4 \kappa} \\
h^- \; = \; & \frac{\rho^2 - \rho \kappa - 4 + 2 \kappa}{4 \kappa}
%h^{\pm} = \frac{1}{4 \kappa} \big( 4 + \rho(\rho+4-\kappa)
%    \pm \sqrt{(\kappa - 4)^2 + 4 \rho (\rho + 4 - \kappa)} \big)
\textrm{.}
\end{align*}

As a %natural
degenerate candidate for producing logarithmic representations, we might
require the above two values to coincide, $h^+ = h^-$. For any $\kappa$
we have a unique way of
satisfying this, the choice $\rho = \frac{\kappa-4}{2}$. In Section
\ref{sec: ell 0} we will see that indeed the local martingales then form
a simple staggered module,
and later we'll see that at $\kappa = 8$ the representation
becomes one very familiar from the physics literature.
%It is also worth noticing that this value
%of $\rho$ has been used for a statement of the duality conjecture
%in \cite{}.

But in fact the requirement of coincidence of $h^\pm$ is not the only way
of getting
coinciding weights for primary fields that arise in the fusion:
a coincidence might happen also with some
descendant field\footnote{This means having a nonvanishing primary descendant
to a primary field, which is not allowed in many conformal field theories.
But in LCFT such is often unavoidable.}.
In such a case we should have $h^+ = h_{r,s}$ for some $r,s \in \bZ_+$ and
$h^- = h^+ + \ell$ where $\ell \geq 0$ is the grade at which the $h^+$
module has a singular vector (or vice versa, the roles of $h^\pm$ exchanged).
For $\kappa \notin \bQ$, the only possibility is $\ell = rs$ and the
desired values of $h^\pm$ can each be obtained by a choice of $\rho$
\begin{align*}
h^+ = h_{r,s} \quad \Leftrightarrow \quad &
    \rho \in \{ \frac{\kappa r - 4 s + \kappa - 8}{2} ,
                \frac{- \kappa r + 4 s + \kappa - 8}{2} \} \\
h^- = h_{r,s} + rs \quad \Leftrightarrow \quad &
    \rho \in \{ \frac{\kappa r + 4 s + \kappa}{2} ,
                \frac{- \kappa r - 4 s + \kappa}{2} \}
\textrm{ .}
\end{align*}
For general $\kappa$ and $r,s$ the two requirements are in conflict with each other.
But for $s=1$ and any $r$ we do obtain solutions at any $\kappa$, namely
$\rho = -2 + \frac{\kappa}{2}(1-r)$ (at specific values of $\kappa$
there may be also other solutions). We remark that requiring
$h^- = h_{r,s}$, $h^+ = h_{r,s} + rs$ instead merely correspods to
changing $\rho$ to $\kappa-4-\rho$.

Apart from the above mentioned case $h^+ = h^-$ we will consider two of the
simplest examples of logarithmic representations that occur at generic
values of $\kappa$. These are the cases $s=1$, $r = 1,2$ above.
In both cases there is a logarithmic coupling $\beta$ to be computed.
In the first case of Section \ref{sec: ell 1} the obtained value of $\beta$
will generalize a proposed infinite series of logarithmic couplings
\cite{MR-logarithmic_M2p} to a continuum of representations here parametrized
by $\kappa$. In the second case, Section \ref{sec: ell 2},
we obtain another formula very similar to such proposals and related
conjectures (although no conjecture about this case had been presented).
At rational values of $\kappa$ further computations
are in principle needed to identify the precise structure of the
representations we encounter.
We will exemplify the phenomenon in Sections
\ref{sec: kappa 8} and \ref{sec: further fusion},
and find again staggered modules familiar from the LCFT literature.

\subsection{A simple calculation for the first staggered module}
\label{sec: ell 0}
As discussed above, $\rho = \frac{\kappa - 4}{2}$ might be a good place
to start looking for logarithms. It is worth mentioning that
this variant was used in the first globally precise formulation of a
duality conjecture for SLEs \cite{Dubedat-duality}.

The partition function $Z \in \Kern A_{\kappa;\rho=(\kappa-4)/2}$ is explicitly
$$Z(x,y) \; = \; (x-y)^{\frac{\kappa - 4}{2 \kappa}} \textrm{ ,}$$
and corresponds to the constant (local) martingale. It is straightforward
to verify with the formulas
of Section \ref{sec: local mgales} that
\begin{align*}
L_n Z =  0 \qquad \textrm{ for $n>0$, and }
\qquad (L_0 - \frac{8-\kappa}{16}) Z  = 0 \textrm{ ,}
\end{align*}
so $Z$ generates a highest weight representation
$\uea Z \subset \Kern A_{\kappa;\rho=(\kappa-4)/2}$ of highest weight
$h_Z = \frac{8-\kappa}{16}$. This weight coincides with
$h_Z = h_{0,1}(\kappa)$. % or with any $h_{a,1 + a \frac{\kappa}{4}}$.
If $\kappa$ is rational the corresponding Verma module
is reducible (chain type), but for $\kappa \notin \bQ$ the
Verma module is irreducible and we readily conclude that
$\uea Z \isom \Verma_{c(\kappa); h_{0,1}(\kappa)}$.

The naive fusion gave the weight $h_Z$ twice denenerate.
This suggests we look not for another singular vector with $L_0$
eigenvalue $h_Z$, but instead for a ``logarithmic partner to $Z$'',
that is a function $\Lambda = \Lambda(x,y;a_2,a_3,\ldots)$
that forms an $L_0$ Jordan block with $Z$
$$ (L_0 - h_Z) \Lambda = Z \textrm{ .}$$
Furthermore the requirement of it being annihilated by positive
modes is in general expected to hold modulo the representation
generated by $Z$. In this case homogeneity leaves us no other 
possibility but $L_n \Lambda = 0$ for $n>0$ which in view of the
explicit formulas for $L_n$ (Section \ref{sec: local mgales})
means that $\Lambda$ should not depend on $a_k$,
$k \geq 2$, and should only be a function of the difference $x-y$.
The obvious candidate
$$ \Lambda(x,y) = (x-y)^{\frac{\kappa-4}{2 \kappa}} \, \log (x-y) $$
satisfies not only the above properties, but most importantly
$A_{\kappa;\rho} \Lambda = 0$. In the language of stochastic processes,
this is the property that
$$\frac{\Lambda(X_t,Y_t)}{Z(X_t,Y_t)} = \log(X_t - Y_t)$$
is a local martingale for the growth process of
$SLE_{\kappa}(\rho=\frac{\kappa-4}{2})$.

In view of the explicit properties above, we have the
following observations about the structure of the representation
$\uea \Lambda \subset \Kern A_{\kappa;\rho=(\kappa-4)/2}$
generated by $\Lambda$. It is indecomposable and contains
the highest weight module $\uea Z$, because $(L_0-h_Z) \, \Lambda = Z$.
The quotient by $\uea Z$ of this module is a highest weight module
of highest weight $h_Z$, as follows from the facts that
$(L_0 - h_Z) \Lambda, L_1 \Lambda, L_2 \Lambda \in \uea Z$.
Therefore $\uea \Lambda$ is in any case a staggered module, with
left and right modules having the same highest weight $h^\lft=h^\rgt=h_Z$.
At irrational $\kappa$
we readily conclude that the staggered module $\uea \Lambda$ fits into
the exact sequence
$$ 0 \longrightarrow \Verma_{c(\kappa), h_{0,1}(\kappa)} \longrightarrow
\uea \Lambda \longrightarrow \Verma_{c(\kappa), h_{0,1}(\kappa)}
\longrightarrow 0 \textrm{ .}$$
From the general theory summarized in Section \ref{sec: staggered theory}
we know that this is enough to identify the
staggered module $\uea \Lambda$.

\subsection{Calculations for staggered modules with $h^\lft \neq h^\rgt$}
\label{sec: ell 1}

\subsubsection{The simplest case}
If the highest weights of the left and right module of a staggered module
are different, the exact sequence alone is not enough to characterize the
module (in general).
The simplest such case is perhaps when the left module
has weight $h_{1,1} = 0$ and there is a logarithmic partner to a
nonvanishing singular vector at grade one. In the discussion we identified
$\rho = -2$ as the choice for which the naive fusion gives highest weights
$h^+=0$ and $h^-=1$.

Indeed, the partition function
$$ Z(x,y) = (x-y)^{-\frac{2}{\kappa}} $$
generates a highest weight representation $\uea Z$ of highest weight
$h_Z = 0 = h_{1,1}(\kappa)$ as follows once we check
\begin{align*}
L_n Z =  0 \qquad \textrm{ for $n>0$, and }
\qquad (L_0 - 0) Z  = 0 \textrm{ .}
\end{align*}
Typically, a highest weight representation with $h=h_{1,1}=0$ in conformal
field theory would only arise as the vacuum representation. While the
Verma module has a singular vector at grade one, there are several
arguments saying that there is none in the vacuum module:
requirement of unitarity alone would suffice, but even relaxing that
the translation invariance of the vacuum should imply the impossibility
of grade one singular vector. But in logarithmic conformal field
theory, the left modules of staggered modules must sometimes
contain nonzero singular vectors, and indeed in the present case we compute that
$$L_{-1} Z \; =\; \frac{4-\kappa}{\kappa} \; (x-y)^{1-\frac{2}{\kappa}}$$
is such a nonzero (if $\kappa \neq 4$) singular vector,
at grade one in $\uea Z$.
The $L_0$ eigenvalue of this singular vector, $h_Z+1=1$ was the other
highest weight $h^-$ we got from the naive computation of fusion in Section
\ref{sec: naive fusion}.
This again suggests looking for a logarithmic partner, now for $L_{-1} Z$.
Let $\Lambda(x,y;a_2,a_3,\ldots)$ denote the function we are looking for.
It should form a $L_0$ Jordan block with $L_{-1} Z$,
$$ (L_0 - 1) \Lambda = L_{-1} Z $$
and positive modes acting on it should give something in the highest
weight submodule $\uea Z$. By considerations of grades,
$L_1 \Lambda \sim Z$ and $L_n \Lambda = 0$ for $n \geq 2$. The latter
implies no dependence on $a_2, a_3, \ldots$ again. A possible solution
to all the equations so far would be
$$ \frac{4-\kappa}{\kappa} \; (x-y)^{1-\frac{2}{\kappa}} \; \log(x-y)
+ b \; (x-y)^{-\frac{2}{\kappa}} \, (x+y) + a \; (x-y)^{1-\frac{2}{\kappa}} $$
although the last term, being already in $\uea \, Z$
will affect none of our computations and can be omitted immediately.
Now in general, dividing this by $Z$ will \emph{not} give a local
martingale of SLE${}_\kappa(\rho=-2)$ --- we see that
$A_{\kappa;\rho} \Lambda$ vanishes only if
$b=\frac{4-\kappa}{8}$, so we define
$$ \Lambda(x,y) = \frac{4-\kappa}{\kappa} (x-y)^{1-\frac{2}{\kappa}} \; \log(x-y)
+ \frac{4-\kappa}{8} \; (x-y)^{-\frac{2}{\kappa}} \, (x+y) \textrm{ .} $$
For irrational $\kappa$ the highest weight modules $\uea Z$ and
$\uea \Lambda / \uea Z$ of respective highest weights $0=h_{1,1}$ and
$1=h_{1,-1}$ must both be Verma modules, so the staggered module
$\uea \Lambda \subset \Kern A_{\kappa; \rho=-2}$ has the exact sequence
$$ 0 \longrightarrow \Verma_{c(\kappa), 0} \longrightarrow
\uea \Lambda \longrightarrow \Verma_{c(\kappa), 1}
\longrightarrow 0 \textrm{ .} $$
The invariant, logarithmic coupling, is then determined by computing
$L_{1} \Lambda = (1-\frac{\kappa}{4}) \, Z$, which gives
$\beta = 1-\frac{\kappa}{4}$.

\bigskip

It is interesting to compare with the study of a consistent operator
algebra for logarithmic CFTs at central charges corresponding to
$\kappa/4 = \frac{p}{2}$, $p =3,5,7,\ldots$,
\cite{MR-logarithmic_M2p}. In that article the authors propose an
operator content of CFT, that would be closed under fusion products.
Based on several computations of fusion products, the authors obtained
that a staggered module with $h^\lft=0$, $h^\rgt=1$ and our
value $\beta = - \frac{p-2}{2}$ of the logarithmic coupling
should be present for all $p$. Admittedly, at these rational values of
$\kappa$ we haven't yet
determined the precise left and right modules, but the value of our
invariant $\beta$ here provides both an interpolation and extrapolation of
the proposal. Here it is moreover obtained by just a few lines of
calculations instead of the use of a complicated algorithm
\cite{Gaberdiel-fusion_in_CFT}
needed to compute the structure of fusion product representations.

\subsubsection{The next example in order of difficulty}
\label{sec: ell 2}

Among the examples proposed in Section \ref{sec: naive fusion},
$h^+ = h_{r,1}$, the case $r=1$ considered above is a lot
easier than the others. In order to show briefly that other cases
are by no means untractable, we record the results of the
next one here.

We take $\rho = \frac{-\kappa-4}{2}$, corresponding to $r=2$.
The partition function $Z(x,y)=(x-y)^{-\frac{1}{2}-\frac{2}{\kappa}}$
generates a highest weight representation of weight $h_Z = h_{2,1}$
\begin{align*}
L_n Z =  0 \qquad \textrm{ for $n>0$, and }
\qquad (L_0 - \frac{3\kappa-8}{16}) Z = 0 \textrm{ .}
\end{align*}
The representation has a non-zero singular vector of conformal weight
$h_{2,1}+2$, 
$$ (L_{-1}^2 - \frac{\kappa}{4} L_{-2}) Z =
    \frac{(\kappa-4)(\kappa-2)(\kappa+4)}{2 \kappa^2} \;
    (x - y)^{\frac{3}{2} - \frac{2}{\kappa}} \textrm{ .}$$
The logarithmic partner that is annihilated by $A_{\kappa;\rho=-(\kappa+4)/2}$
can be chosen as
\begin{align*}
\Lambda(x,y) \; = \;& \frac{(\kappa-4)(\kappa-2)(\kappa+4)}{2 \kappa^2} \Big(
    (x-y)^{\frac{3\kappa-4}{2\kappa}} \log(x-y) \\
& \qquad + (x-y)^{\frac{-\kappa-4}{2\kappa}}
    \; \frac{4 \kappa \, x^2 - (4\kappa + \kappa^2) \, y^2 }{16} \Big)
\end{align*}
and with this normalization we have 
\begin{align*}
(L_0 - \frac{3\kappa + 24}{16}) \Lambda =
    (L_{-1}^2 - \frac{\kappa}{4} L_{-2}) Z \quad , & \quad
    L_n \Lambda \in \uea Z \; \textrm{ for $n>0$.}
\end{align*}
Finally we compute $(L_1^2 - \frac{\kappa}{4} L_2) \Lambda = \beta Z$
with $\beta = -\frac{1}{16} (\kappa-4)(\kappa-2)(\kappa+4)$, which
is a formula very similar to the conjectured and proposed values of
logarithmic couplings in
\cite{MR-logarithmic_M2p}\footnote{The corresponding fusion wasn't
studied in the article. In a private communication David Ridout confirmed
that it can nevertheless be treated with the same methods and
that the values of $\beta$ agree with the one found here.}.

\subsection{A remark on contragredients}
\label{sec: contragredients}

Before going on to other examples, which we will compare with fusion product
representations\footnote{Constructing a representation of Virasoro algebra
associated to a fusion (more precisely, to two representations to be fused)
as described in \cite{Gaberdiel-fusion_in_CFT},
is a much more satisfactory definition of fusion in CFT than the naive
connsiderations of Section \ref{sec: naive fusion}. We shall not go to the
details of this construction, but merely compare our results with fusion
products computed in the literature.} denoted in what follows by $\fprod$,
we pause briefly to comment on what we believe in general to
be the relation between local martingales and fusions.
We will often find that local martingales fall into representations
that coincide with the fusion product representation. This is not quite the
exact statement that we'd like to make, though.
Already the example of chordal SLE${}_\kappa$ in
Sections \ref{sec: chordal SLE representation} and \ref{sec: Cardys field},
as well as the one of Section \ref{sec: first fusion},
indicate that local martingales form not the fusion product module
of all boundary changing fields, but rather its contragredient.

In this light, any observed coincidence of certain fusion
product and the space of local martingales should be supplemented
by the fact that the module in question is contragredient
to itself.\footnote{This is analogous to the phenomenon with
chordal SLE with irrational $\kappa$ where we had
$\Kern A_{\kappa} \isom \sQ_{1,2}^* \isom \sQ_{1,2} \isom \sL_{h_{1,2}(\kappa)}$.}
Here we will check that this is the case with the examples above at
irrational $\kappa$, and in
later examples we will content ourselves to remark what needs to be changed
in the arguments.

For the case of Section \ref{sec: ell 0}, we choose from the dual of
$(\uea \Lambda)_{[h_Z]} = \bC Z \oplus \bC \Lambda$ two elements
$\xi^*,\eta^*$ such that $\bra \xi^*, \Lambda \ket = 1 = \bra \eta^*, Z \ket$
and $\bra \xi^*, Z \ket = 0 = \bra \eta^*, \Lambda \ket$.
Now by the formula defining
contragredients, $\bra (L_0-h_Z) \eta^*, \Lambda \ket =
\bra \eta^*, (L_0-h_Z) \Lambda \ket = \bra \eta^*, Z \ket = 1$
whereas all other pairings of $(L_0-h_Z) \eta^*$ and
$(L_0-h_Z) \xi^*$ with $Z$ and $\Lambda$ vanish. Therefore
$(L_0-h_Z) \eta^* = \xi^*$ and $(L_0-h_Z) \xi^* = 0$.
By an obvious consideration of grades, positive
generators must annihilate both vectors,
$L_n \xi^* = 0 = L_n \eta^*$ for $n>0$. Thus $\uea \xi^*$ is a highest
weight module of highest weight $h_Z=h_{0,1}(\kappa)$,
at irrational $\kappa$ therefore the Verma module.
The same applies to $\uea \eta^* / \uea \xi^*$.
By observing that the graded dimensions of $\uea \Lambda$ and $\uea \eta^*$
must now be equal,
the full contragredient module $(\uea \Lambda)^*$ is seen to be a staggered
module with the exact sequence
$$ 0 \longrightarrow \Verma_{c(\kappa), h_{0,1}(\kappa)} \longrightarrow
(\uea \Lambda)^* \longrightarrow \Verma_{c(\kappa), h_{0,1}(\kappa)}
\longrightarrow 0 \textrm{ ,} $$
which in this case identifies the module, and proves that
$\uea \Lambda \isom (\uea \Lambda)^*$.

For the case of Section \ref{sec: ell 1} one proceeds very similarly.
Choose $\xi^*$ in the one-dimensional space $(\uea \Lambda)^*_{[0]}$ such that
$\bra \xi^*, Z \ket=1$.
This $\xi^*$ then generates a highest weight submodule $\uea \xi^*$
of the contragredient $(\uea \Lambda)^*$, with highest weight $0$.
Now $L_{-1} \xi^*$ doesn't couple to the singular
vector $L_{-1} Z$, but still it is nonzero because
$\bra L_{-1} \xi^* , \Lambda \ket = \bra \xi^* , L_1 \Lambda \ket =
\beta \neq 0$ (when $\kappa \neq 4$). We can therefore choose a vector
$\eta^* \in (\uea \Lambda)^*_{[1]}$ that together with $L_{-1} \xi^*$
spans the grade $1$ of the contragredient, for example normalized by
$\bra \eta^* , a \Lambda + b L_{-1} Z \ket = b \beta$. With
$\bra (L_0 - 1) \eta^*, a \Lambda + b L_{-1} Z \ket =
\bra \eta^* , a L_{-1} Z \ket = a \beta$ we see that
$(L_0-1) \eta^* = L_{-1} \xi^*$. A final reference to the
graded dimension shows that for $\kappa \notin \bQ$ the contragredient
$(\uea \Lambda)^*$ is a staggered module with the same exact sequence
$$ 0 \longrightarrow \Verma_{c(\kappa), 0} \longrightarrow
(\uea \Lambda)^* \longrightarrow \Verma_{c(\kappa), 1}
\longrightarrow 0$$
and its invariant
$\bra L_1 \eta^*, Z \ket = \bra \eta^*, L_{-1} Z \ket = \beta$
is the same, so $(\uea \Lambda)^* \isom \uea \Lambda$ again.

Since our main concern is local martingales we will mostly omit detailed
discussions about contragredients in what follows.

\subsection{Specializations to $c=-2$}
\label{sec: kappa 8}

The conformal field theories of central charge $c=-2$ are perhaps the
most studied examples of logarithmic conformal field theory, indeed
already discussed in the article that introduced the concept of logarithmic
operators \cite{Gurarie-logarithmic_operators}.
The merits of $c=-2$ are at least partly explained by the fact that all
reducible Verma modules are of the chain type, % ($\mathrm{III}^{0}$)
but also the existence of free fermionic theories (symplectic fermions
\cite{Kausch-symplectic_fermions})
has helped to explore this case.

We will therefore compare the results of the previous sections to the CFT
literature, by choosing appropriate values of $\kappa$ of course.
Below we specialize the first two examples to $\kappa=8$ and observe
coincidence with known results from fusion products.

We also mention that the chordal SLE${}_{\kappa=8}$ has been shown to
be the continuum limit of uniform spanning tree Peano curve
\cite{LSW-LERW_and_UST},
which is a discrete curve passing through
all lattice points while surrounding the (uniformly chosen random
spanning) tree.
It could have a physical interpretation as a single space filling polymer.

\subsubsection{The module of Section \ref{sec: ell 0} at $\kappa=8$}

From the literature of fusion products in CFT, the simplest way to
produce a logarithmic representation through
fusions seems to be to consider $c=-2$ and the fusion of two
$\sQ_{\kappa=8;1,2} = \sL_{c=-2; h=-\frac{1}{8}}$
representations %($h_{1,2}(8) = -\frac{1}{8}$)
\cite{GK-indecomposable_fusion_products, %Kausch-symplectic_fermions,
Gurarie-logarithmic_operators}.
Fortunately for us, the field at the tip
of the SLE${}_{\kappa=8}$ trace has weight $h_{1,2}(\kappa=8)=-\frac{1}{8}$,
and if we take $\rho = 2 = \frac{\kappa-4}{2}$
as in Section \ref{sec: ell 0}, the field at the marked point has the
same weight.

\begin{figure}
\begin{center}
\includegraphics[width=0.8\textwidth]{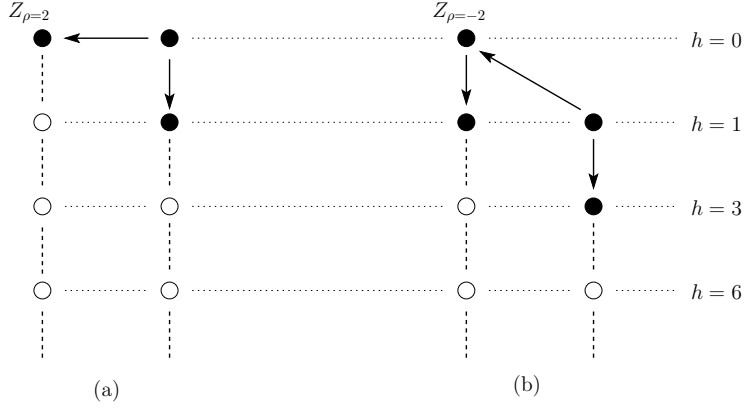}
\caption{
(a) \emph{Local martingales of SLE${}_{\kappa=8}(\rho=2)$ form a
simple staggered module which coincides with the fusion
$\sQ_{1,2} \fprod \sQ_{1,2}$.}
(b) \emph{For SLE${}_{\kappa=8}(\rho=-2)$ we have another staggered
module. It now coincides with the fusion $\sQ_{1,2} \fprod \sQ_{2,2}$.
We point out that to identify the module in this picture,
we had to determine also the value of an invariant $\beta$.}}
\label{fig: SLE8 modules}
\end{center}
\end{figure}

We have already seen that $\Lambda$ generates a staggered module
$\uea \, \Lambda \subset \Kern A_{\kappa=8; \rho=(\kappa-4)/2}$ of
local martingales, with the left module generated by
the partition function $Z$
\begin{align*}
0 \longrightarrow \sU \, Z \longrightarrow \sU \, \Lambda
      \longrightarrow (\sU \, \Lambda) / (\sU \, Z) \longrightarrow 0
\textrm{ .}
\end{align*}
To decide the structures of the left and right modules, it suffices to compute
the following graded dimensions merely up to grade three
(with the help of a computer the task is easy so we checked it up to grade six) 
\begin{align*}
\chi_{\sU \, \Lambda} (q) \; & = \; \sum_{m \geq 0}
    \big( \dmn (\uea \Lambda)_{[m]} \big) q^m \\
= \; & 2 + 1 q + 3 q^2 + 3 q^3 + 6 q^4 + 7 q^5 + 12 q^6 + \cdots
\quad \textrm{ and } \\
\chi_{\sU \, Z} (q) \; & = \; \sum_{m \geq 0}
    \big(\dmn (\uea Z)_{[m]} \big) q^m \\
 = \; & 1 + q^2 + q^3 + 2 q^4 + 2 q^5 + 4 q^6 + \cdots = \chi_{\sQ_{1,1}}(q)
\textrm{ .}
\end{align*}
These imply that the left module is irreducible\footnote{In this case
the computation of the structure of $\uea \, Z$ is still easy to do by
hand, and we present an alternative argument that was also important for
\cite{KK-reversibility_duality}.
The check that the partition function $Z (x,y) = (x-y)^{1/4}$ of
SLE${}_{\kappa=8}(\rho=2)$ generates a highest weight module of weight
$h=0$ is done as usual by computing $L_n Z = 0$ for $n>0$ and
$(L_0 - 0) Z = 0$ with the formulas of Section \ref{sec: local mgales}.
Then we observe a coincidence at $\kappa=8$
of $\rho=2=\frac{\kappa-4}{2}$ and $\rho=\kappa-6$, so the partition
function is in fact also $Z(x,y) = (x-y)^{-h_{1,2}}$. Now doing the easy
computation
\begin{align*}
L_{-1} \, Z (x,y)
\; = \; & \Big( (x^2 - 3 a_2) \pder{x} + (y^2 - 3 a_2) \pder{y}
    + 2 h_{1,2} (x+y) \Big) (x-y)^{-2 h_{1,2}} \; = \; 0
\textrm{ ,}
\end{align*}
and recalling that the Verma module $\Verma_{c=-2;h=0}$
has maximal submodule generated by $L_{-1} \, v_{c=-2;h=0}$,
we readily conclude that $\sU \, Z$ is irreducible.}
$\uea \, Z \isom \sL_{h=0} = \sQ_{1,1}$, and that
the quotient $\sU \, \Lambda / \sU \, Z$, which is a
highest weight module, has graded dimension
$\chi_{\sU \, \Lambda} (q) - \chi_{\sU \, Z} (q) = \chi_{\sQ_{1,3}}(q)$.
In another words the structure of $\sU \, \Lambda$ can be read from
the exact sequence
\begin{align*}
0 \longrightarrow \sQ_{\kappa=8;1,1} \longrightarrow \sU \, \Lambda
      \longrightarrow \sQ_{\kappa=8;1,3} \longrightarrow 0
\end{align*}
or pictorially from Figure \ref{fig: SLE8 modules}(a). Therefore, the module
$\uea \Lambda \subset \Kern A_{\kappa=8; \rho=2}$ is indeed
isomorphic to the fusion $\sQ_{\kappa=8;1,2} \fprod \sQ_{\kappa=8;1,2}$
computed in \cite{GK-indecomposable_fusion_products}.

As in Section \ref{sec: contragredients}, it is easy to show that this module
is contragredient to itself: indeed the only thing that has to be added to
the argument is the observation that now $L_{-1} \xi^* = 0$ but
$L_{-1} \eta^* \neq 0$, after which graded dimensions imply
$\uea \xi^* \isom \sQ_{1,1} \isom \uea Z$ and
$\uea \eta^* / \uea \xi^* \isom \sQ_{1,3} \isom \uea \Lambda / \uea Z$.
In conclusion, we have
$$(\sQ_{\kappa=8;1,2} \fprod \sQ_{\kappa=8;1,2})^* \isom \uea \Lambda$$
so we have found local martingales that form a module contragredient
to the appropriate fusion product.

\bigskip

A probabilistic interpretation of the SLE variant used here is also
readily available. Indeed, it has been proved that chordal
SLE${}_{\kappa=8}$ is the the scaling limit of uniform spanning tree
Peano curve \cite{LSW-LERW_and_UST}. But the chordal SLE
in $\bH$ from $X_0$ to $Y_0$ (obtained by a M\"obius coordinate
change from the one towards infinity, Section \ref{sec: SLE variants})
is an SLE${}_\kappa(\rho=\kappa-6)$.
Therefore the variant considered above
is nothing but the scaling limit of a uniform spanning tree Peano curve,
in $\bH$ from $X_0$ to $Y_0$.
The fact that $Y_0$ now is the endpoint
of the curve also explains the coincidence of weights
of the two fields, $h_x=h_y=h_{1,2}$.

\subsubsection{Module of Section \ref{sec: ell 1} at $\kappa=8$}
\label{sec: kappa 8 ell 1}

A fusion that is almost as simple as the above is
$\sQ_{\kappa=8;1,2} \fprod \sQ_{\kappa=8;2,2}$
\cite{GK-indecomposable_fusion_products}.
The weights of the fields to be fused are those of Section \ref{sec: ell 1},
$h_x = h_{1,2} = \frac{-1}{8}$ and $h_y = h_{2,2} = \frac{3}{8}$, so
we already know that there's a staggered module of local martingales
of SLE${}_{\kappa=8}(\rho=-2)$.
Because the Verma modules of weights $h^+=0=h_{1,1}$ and $h^-=1=h_{2,1}$
are of chain type, we still need to identify the
left and right modules, which is easiest done by calculations of graded
dimensions
\begin{align*}
\chi_{\uea \Lambda} (q) \; = \; & \sum_{m \geq 0}
    \big( \dmn (\uea \Lambda)_{[m]} \big) q^m \\
= \; & 1 + 2q + 3q^2 + 4q^3 + 7q^4 + 10q^5 + 14q^6 + \cdots \\
\chi_{\uea Z} (q) \;= \; & \sum_{m \geq 0}
    \big( \dmn (\uea Z)_{[m]} \big) q^m \\
= \; & 1 + 1q + 2q^2 + 2q^3 + 4q^4 + 5q^5 + 8q^6 + \cdots
\end{align*}
from which we conclude that $\uea \Lambda \subset \Kern A_{\kappa=8;\rho=-2}$
has the exact sequence
\begin{align*}
0 \longrightarrow \sQ_{\kappa=8;1,3} \longrightarrow \sU \, \Lambda
      \longrightarrow \sQ_{\kappa=8;1,5} \longrightarrow 0
\textrm{ .}
\end{align*}
Both the left and right modules and the invariant $\beta = -1$
(computed in Section \ref{sec: ell 1})
again agree with results of the fusion products
\cite{GK-indecomposable_fusion_products}, and the module is contragredient
to itself. Figure \ref{fig: SLE8 modules}(b) illustrates the structure.

%  *****************************************************************
%  *  REPRESENTATIONS  *********************************************
%  *****************************************************************
\section{Fusions in percolation with SLE${}_6$ variants}
\label{sec: kappa 6}

In this section we consider in some more detail the case $\kappa=6$,
which is related to critical percolation. When explicit
reference to $\kappa$ is omitted in the rest of the section, this
is the value meant.

Before turning to our concrete examples in this case, we recall a few
particular features of the conformal field theory approaches to
percolation. First, it has been accepted for a long time that
critical percolation should be described by
a conformal field theory of central charge $c=0$. Indeed, one definite
success of conformal field theories was Cardy's exact formula for the
crossing probability of a conformal
quadrilateral, derived originally from CFT \cite{Cardy-percolation},
immediately supported by numerical evidence
\cite{LPPS-crossing_probabilities_in_percolation} and later proved
rigorously for triangular lattice site percolation
\cite{Smirnov-critical_percolation}.
On the other hand, the $c=0$ minimal model would be a trivial theory
containing nothing but the identity operator, and as such clearly
inadequate. There are no obvious alternative candidates, and for
this reason the question of operator content of
percolation has received quite a lot of attention recently
\cite{MR-percolation_LCFT, RP-fusion_algebra_of_percolation,
EF-fusion_for_augmented_minimal_models, RS-associative_algebraic_LCFT}.
We will keep these recent studies in mind while exploring the modules
of local martingales and their relation to fusion products.

First, to make precise in which sense SLE${}_{\kappa=6}$ describes
percolation, the discrete site percolation model and result about
the scaling limit is recalled in Section \ref{sec: exploration path}.
Having the discrete case in mind allows to interpret the SLE variants
in terms of percolation events, which also facilitates comparisons to
CFT literature.

We will argue that the module $\Kern A$ for ordinary chordal SLE reflects
the boundary field of Cardy's crossing probability argument,
in Section \ref{sec: Cardys field},
and for that we exhibit more carefully the degeneracies that even this
simplest variant has at rational values. A very simple fusion without
logarithms is considered in Section \ref{sec: first fusion},
again in some detail. There the lesson to take home is that in the space of
local martingales, not all null field conditions have yet been taken
into account, only the one corresponding to the field at the tip of the
curve. Section \ref{sec: further fusion} exhibits the simplest
way to produce logarithms in percolation CFT. The variant used admits
a crossing event interpretation, and the module of local martingales is
once again in agreement with CFT fusion prediction. The final example
of Section \ref{sec: existence of Q21} is
of somewhat speculative nature: we consider a variant of
SLE${}_{\kappa=6}$ whose local martingales form a module that has been
argued to be inconsistent with the percolation
CFT, and discuss whether the variant could be made appear in a natural
probabilistic setting in percolation.

\subsection{The percolation exploration path}
\label{sec: exploration path}

\begin{figure}
\begin{center}
\includegraphics[width=1.0\textwidth]{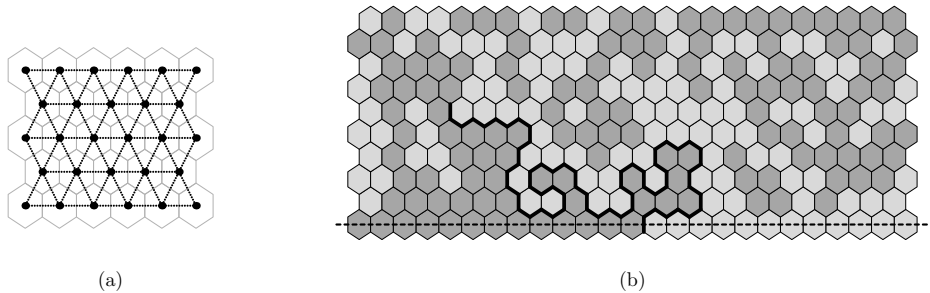}
\caption{
(a) \emph{The sites of triangular lattice are the centers of the faces
of hexagonal lattice.}
(b) \emph{The exploration path of 
percolation separates a ``black'' cluster from a ``white'' one.
Boundary conditions are ``black'' on the left and ``white''
on the right.
%In conformal field theory the change in boundary conditions
%corresponds to insertion of the Cardy's operator $\psi$.
}}
\label{fig: exploration}
\end{center}
\end{figure}

A configuration of site percolation on the infinite triangular lattice
$T$ is $\sigma \in \{ -1, +1\}^T$ which is interpreted as
a coloring of sites $z \in T$ as black ($\sigma_z = -1$) or
white ($\sigma_z = +1$).
%an association
%of colors black ($-1$) or white ($+1$) to all sites of the lattice
%$z \in T$.
It is convenient to think of the sites of the triangular
lattice as the faces of the dual, hexagonal lattice $H$, see
Figure \ref{fig: exploration}(a). For $p \in [0,1]$ the percolation
measure with parameter $p$ is the Bernoulli probability measure with
parameter $p$ on the space of configurations $\{-1, +1 \}^T$, that is
we choose the color of each site (hexagon) independently to be white
with probability $p$ and black with probability $1-p$. We consider the
model at its critical point $p = p_c = 1/2$.

The continuum limit corresponds to taking the lattice spacing to zero.
Consider therefore the triangular lattice with lattice spacing $\delta$,
denoted by $\delta T$. Let $D^\delta \subset \delta T$ be a lattice domain,
simply connected in the lattice sense and such that the lattice boundary
$\bdry D^\delta$, consisting of sites adjacent to those of the the domain,
is a simple lattice path. We imagine splitting the boundary
$\bdry D^\delta$ to two complementary subarcs
$\overline{b^\delta a^\delta}$ and $\overline{a^\delta b^\delta}$,
where $a^\delta$ and $b^\delta$ are lattice edges ($b^\delta$ can be
allowed to be at infinity). Consider the percolation configuration
restricted to $D^\delta$ and extended by black on
$\overline{b^\delta a^\delta}$ and white on $\overline{a^\delta b^\delta}$.
The exploration path of percolation in $D^\delta$ from $a^\delta$ to
$b^\delta$ is the maximal dual lattice path $\hat{\gamma}^\delta$
starting from $a^\delta$ such that each edge of $\hat{\gamma}^\delta$
is adjacent to one black and one white site of the above configuration.
Figure \ref{fig: exploration}(b) portrays a percolation configuration
and the beginning of the corresponding exploration path in
$D^\delta = \bH \cap \delta T$ with $b^\delta$ at infinity.

At the critical point, these random paths have a limit for example in
the sense of weak convergence of the probability laws on the space of
unparametrized paths with the metric
\begin{align*}
d(\gamma^1, \gamma^2) \; = \;
\underset{\substack{\textrm{parametrizations} \\
    \gamma^j = \tilde{\gamma}^j [0,1] % \quad \textrm{ for $j=1,2$}
        }}{\inf }
\; \sup_{s \in [0,1]} |\tilde{\gamma}^1 (s) - \tilde{\gamma}^2 (s)|
\textrm{ .}
\end{align*}
Let $D$ be a Jordan domain (simply connected bounded open set of the
plane whose boundary is a simple closed curve) with two distinct boundary
points $a,b \in \bdry D$. In
\cite{Smirnov-critical_percolation, CN-critical_percolation_exploration_path}
it is shown that if one approximates $(D,a,b)$ in a suitable sense by
lattice domains $(D^\delta, a^\delta, b^\delta)$, then the laws of
the exploration paths $\hat{\gamma}^\delta$ of critical percolation
in $D^\delta$ from $a^\delta$ to $b^\delta$ converge to the image
$f_D(\gamma)$ of chordal SLE${}_6$ trace $\gamma$ in $\bH$ under any
conformal map
$f_D : \bH \rightarrow D$ such that $f_D(0) = a$, $f_D(\infty) = b$.
%Below we thus sometimes call the SLE${}_6$ trace $\gamma$ the (continuum
%limit of the critical) percolation exploration path.

\bigskip

The convergence results for exploration path have proved powerful. They
have so far been used at least to the mathematical treatments of critical
exponents \cite{SW-critical_exponents_for_percolation}, one-arm exponent
\cite{LSW-one_arm_exponent_for_percolation}, Watts' crossing
formula \cite{Dubedat-excursion_decomposition} and surrounding
probabilities \cite{Schramm-percolation_formula}.

\subsection{Chordal SLE${}_6$ and Cardy's boundary field}
\label{sec: Cardys field}

The result described above allows us to interpret the chordal
SLE${}_{\kappa=6}$ as the scaling limit of the exploration path as in Figure
\ref{fig: exploration}(b).

We have already considered the module of local martingales in the
case of chordal SLE${}_\kappa$,
which for irrational $\kappa$ was an irreducible highest weight module
with highest weight vector $Z=1$. As mentioned in Section
\ref{sec: chordal SLE representation}, in general we have
$\Kern A_{\kappa} \isom \sQ_{1,2}^*$.
Below we take a closer look at $\kappa=6$ to illustrate the degeneracies
that occur at rational values of $\kappa$. There is another important reason
to discuss this case: this simple variant which doesn't yet
involve fusions will still teach us something, it will exhibit a representation
contragredient to the conformal family of the boundary condition changing
field $\psi$ used by Cardy in the derivation of the 
crossing probability formula.

In our case $\kappa=6$, $h_{1,2} = 0 = h_{1,1}$ and $\sQ_{1,2}$ contains
a non-zero singular vector at level $1$, $L_{-1} \, w_{1,2} \neq 0$.
At level $2$ there is a null vector,
$(L_{-1}^2 - \frac{2}{3} \, L_{-2} ) \, w_{1,2} = 0$.
These properties correspond
exactly to what one has to assume of Cardy's boundary changing
field $\psi$ (see \cite{MR-percolation_LCFT}).
It should be a primary field with weight $h = 0$ and have
a null descendant at level two (leading to a second order differential
equation for correlation functions), but it should not be the identity,
$L_{-1} \, w_{1,2} \neq 0$ so that $\psi$ is not translation
invariant (Cardy's crossing probability formula is non-constant!).

As $\sQ_{1,2}$ is reducible at $\kappa=6$, its
contragredient $\sQ_{1,2}^*$ is not a highest weight representation.
The contragredient contains a subrepresentation isomorphic to the
irreducible module $\sL_{h=0} = \sU \, w_{1,2}^* \, \subset \, \sQ_{1,2}^*$
generated by $w^*_{1,2} \in (\sQ^*_{1,2})_0$ such that
$\bra w_{1,2}^* , w_{1,2} \ket = 1$
(in this case $\sL_{h=0}$ happens to be one dimensional).
The full contragredient module %$\sQ^*_{1,2}$
is generated by a sub-singular vector\footnote{We use the term sub-singular
here to describe a vector that is a representative of a singular vector
in a quotient by submodule, here in $\sQ_{1,2}^* / \sL_{h=0}$.}
$\xi^* \in (\sQ_{1,2}^*)_1$ that has a nonzero value on the singular vector,
$\bra \xi^*, L_{-1} \, w_{1,2} \ket = 1$. We have
$\sQ_{1,2}^* / \sL_{h=0} \isom \sL_{h=1}$, or in other words
$$ 0 \longrightarrow \sL_{h=0} \longrightarrow \sQ_{1,2}^* \longrightarrow
\sL_{h=1} \longrightarrow 0 \textrm{ ,}$$
but of course $\sQ_{1,2}^*$ is not a staggered module: $L_0$ is still
diagonalizable!

\begin{figure}
\begin{center}
\includegraphics[width=0.75\textwidth]{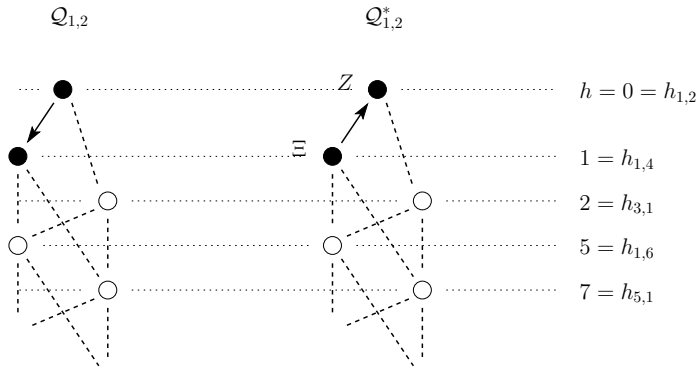}
\caption{\emph{Cardy's boundary changing field $\psi$ generates a
conformal family isomorphic to $\sQ_{1,2}$.
The chordal SLE${}_{\kappa=6}$ local martingales form the contragredient module
$\sQ_{1,2}^*$.
}}
\label{fig: Q12 module}
\end{center}
\end{figure}

To see $\sQ_{1,2}^*$ arise
as module of SLE local martingales is very easy:
the one dimensional module $\sL_{h=0}$ is generated by the partition
function $Z=1$. To verify that $\sU \, Z$ is the irreducible highest
weight module we only check
\begin{align*}
L_{2} \; Z = L_{1} \; Z = L_{0} \; Z
= L_{-1} \; Z = L_{-2} \; Z \; = \; 0
\textrm{ .}
\end{align*}
The function representing $\xi^*$ is $\Xi(x,a_2,a_3,\ldots) = x$,
and the fact that $\Xi \in \Kern A$ is just the observation that
$X_t = \sqrt{6} \; B_t$ is a (local) martingale. By the above general arguments
we expect that $(\sU \, \Xi) / (\sU \, Z) \isom \sL_{h=1}$.
After verifying that $L_1 \, \Xi = 1 = Z$, $L_2 \, \Xi = 0$
and $(L_0 - 1) \, \Xi = 0$, i.e.
that $\Xi$ becomes singular in the quotient, there are several ways to
determine the explicit structure of the quotient 
$(\sU \, \Xi)$ by $(\sU \, Z)$, which must now be a highest weight module.
The result is expected, %(as it should) that
$(\sU \, \Xi) / (\sU \, Z) \isom \sL_{h=1}$,
which is most concretely verified by showing (with the help of a computer)
that there are null vectors at the levels $4$ and $6$
(note that $h=1=h_{1,4}=h_{3,2}$)
\begin{align*}
& \chi_{1,4} \, \Xi \; = \; 0 \quad \textrm{ and } \quad
    \chi_{3,2} \, \Xi \; = \; 0 \quad \textrm{ , where} \\
\chi_{1,4} \; = \; & L_{-1}^4 - \frac{20}{3} \, L_{-2} L_{-1}^2 + 4 \, L_{-2}^2
    + 4 \, L_{-3} L_{-1} - 4 \, L_{-4} \\
\chi_{3,2} \; = \; & L_{-1}^6 - 14 \, L_{-2} L_{-1}^4
    + \frac{112}{3} \, L_{-2}^2 L_{-1}^2
    - \frac{512}{27} \, L_{-2}^3 \\
& + 14 \, L_{-3} L_{-1}^3 - \frac{40}{3} \, L_{-3} L_{-2} L_{-1}
    - \frac{208}{9} \, L_{-3}^2 - 48 \, L_{-4} L_{-1}^2 \\
& + \frac{688}{9} \, L_{-4} L_{-2} + \frac{88}{9} \, L_{-5} L_{-1}
    + \frac{80}{3} \, L_{-6}
\textrm{ .}
\end{align*}
The same conclusion is obtained slightly less explicitly by computing the
number of linearly independent functions among
$L_{-n_1} \cdots L_{-n_m} \, \Xi$, with
$n_1 \geq n_2 \geq \cdots \geq n_m > 0$ and $\sum_j n_j = k \geq 0$.
Since $\sU_{k+1} \, Z = \{ 0 \}$, this gives the graded dimension of
$(\sU \, \Xi) / (\sU \, Z)$
\begin{align*}
\chi_{\sU^- \, \Xi}(q)
\; = \; & \sum_{k \geq 0} \big( \dmn (\sU \Xi)_{[1+k]} \big) \; q^k \\
\; = \; & 1 + q + 2 q^2 + 3 q^3 + 4 q^4 + 6 q^5 + 8 q^6 + \cdots \\
\; = \; & p(0) + p(1) q + p(2) q^2 + p(3) q^3 + (p(4)-p(0)) q^4 \\
& \qquad + (p(5)-p(1)) q^5 + (p(6)-p(2)-p(0)) q^6 + \cdots
\textrm{ .}
\end{align*}
Yet an alternative argument for which no further calculations are needed
relies on the character identity
$\chi_{\sL_{h=0}} (q) + q \; \chi_{\sL_{h=1}} (q)
= (1-q^2) \; \prod_{j=0}^\infty (1-q^j)^{-1} = \chi_{\sQ_{1,2}} (q)$.
From it we read that if $(\sU \, \Xi) / (\sU \, Z)$ is irreducible,
the graded dimension of $\sU \, \Xi$ is that of
$\Kern A \subset \sF^\poly$ (for which we use the result of
\cite{BB-SLE_martingales}). If $\chi_{1,4} \, \Xi$ or $\chi_{3,2} \, \Xi$
would not belong to $\sU \, Z$ so that the quotient would be reducible,
the dimensions would be too large.
The structure of the modules is visualized in Figure \ref{fig: Q12 module}.

\bigskip

This case of chordal SLE${}_\kappa$ is well understood and not new.
But at $\kappa = 6$ we've seen that it is
related to Cardy's field $\psi$ and can be thought of a
concrete verification that the representation $\sQ_{1,2}$ appears
in percolation.
It has been understood already before that Cardy's field $\psi$
is the boundary condition changing field that changes between fixed black
and fixed white boundary conditions, as Figure \ref{fig: exploration}(b)
also illustrates.

%To proceed beyond this easiest case, we'll have to be able to address
%fusions.

\subsection{SLE${}_6(0)$ and fusion of two Cardy's fields}
\label{sec: first fusion}

After arguing the existence of $\sQ_{1,2}$, if the CFT is to be closed under
fusion, a natural thing to do is to
proceed by taking fusion products of this representation as in
\cite{MR-percolation_LCFT}.
As a first step one
considers $\sQ_{1,2} \, \fprod \, \sQ_{1,2}$,
the result of which is
$\sQ_{1,1} \oplus \sL_{h=1/3}$,
\cite{MR-percolation_LCFT, EF-fusion_for_augmented_minimal_models,
RS-associative_algebraic_LCFT, RP-fusion_algebra_of_percolation}.

To study this fusion, we will find an SLE variant where another
boundary changing field $\psi$ is present in addition to that
at $X_0$. In fact, the endpoint (above chosen to be $\infty$) of the
curve plays the same role in boundary conditions as the starting
point $X_0$ and thus in a chordal SLE from $X_0 = 0$ to $Y_0 \neq 0$
the boundary changing fields on the real axis should provide the two
modules we want to fuse.
The chordal SLE towards $Y_0$ is obtained by a coordinate change
as discussed in Section \ref{sec: SLE variants}, and is in general an
SLE${}_\kappa(\rho)$ with $\rho = \kappa-6$. In our case
$\kappa = 6$ and $\rho = 0$ so again the partition function is $Z=1$
and statistics of the driving process are not affected
(up to times until which the variant SLE${}_6(0)$ is defined).
Actually, this curiosity is easily interpreted probabilistically:
it is a special case of a property known as the locality of SLE${}_6$.
Notice that whether the exploration path takes a left or right turn is
determined by the color of hexagon in front of the path, which is chosen
independently of other colors. Thus the statistics of the path (up to
the time it reaches the changed part of the boundary)
is not affected by whether the boundary colors change at
$\infty$ or $Y_0$.

\subsubsection{Local martingales for SLE${}_6(0)$}
We now consider the process SLE${}_6(0)$ and the functions
$\varphi(x,y,a_2,a_3,\ldots)$ such that
$\frac{1}{Z} \, \varphi(X_t,g_t(Y_0),a_2(t),\ldots)$ are local martingales.
If $\varphi$ doesn't depend on $y$ the question reduces to the one
in Section \ref{sec: chordal SLE representation}. In particular
the representations found there are contained in $\Kern A_{\rho=0}$.

Since the fusion $\sQ_{1,2} \, \fprod \, \sQ_{1,2}$ splits into
a direct sum of representations where $L_0$ eigenvalues are in
$0 + \bN$ and $\frac{1}{3} + \bN$, the contragredient splits accordingly.
We start by discussing the former part, $\sQ_{1,1}^*$.
In Section \ref{sec: chordal SLE representation} we have already
found $Z$ and $\Xi$ such that
$\sL_{h=0} \isom \sU \, Z \subset \sU \, \Xi
\subset \Kern A$. In $\sQ_{1,1}$ there's a non-zero singular vector
at level $2$ and in $\sQ_{1,1}^*$ one correspondingly has a
sub-singular vector $\theta^* \in (\sQ_{1,1}^*)_2$.
This generates the full representation $\sU \, \theta^* = \sQ_{1,1}^*$
and the quotient is $(\sU \, \theta^*) / (\sU \, w_{1,1}^*)
\isom \sL_{h=2}$. With a little imagination or familiarity with Bessel
processes, one finds the corresponding
local martingale for SLE, $\frac{1}{5} (X_t-Y_t)^2 - a_2(t)$. Denote
$\Theta(x,y,a_2,a_3,\ldots) = \frac{1}{5} (x-y)^2 - a_2 \in \sF$. One
explicitly checks that
\begin{align*}
L_1 \, \Theta = 0 \qquad L_2 \; \Theta = 1 = Z \qquad &
(L_0 - 2) \, \Theta = 0
\end{align*}
so that $(\sU \; \Theta) / (\sU \; Z)$ is a highest weight
module of weight $h=2=h_{3,1}=h_{1,5}$ and
\begin{align*}
\big( L_{-1}^3 - 6 \; L_{-2} L_{-1} + 6 \; L_{-3} \big)
    \; \Theta & \; = \; 0 \\
\big( L_{-1}^5 - \frac{40}{3} L_{-2} L_{-1}^3
+ \frac{256}{9} L_{-2}^2 L_{-1} + \frac{52}{3} L_{-3}L_{-1}^2
\qquad \qquad & \\
- \frac{256}{9} L_{-3}L_{-2} - \frac{104}{3} L_{-4}L_{-1}
+ \frac{208}{9} L_{-5} \big) \; \Theta & \; = \; 0
\end{align*}
so that there are null vectors at
levels $3$ and $5$. We conclude that the module $\sU \; \Theta$ is indeed
isomorphic to $\sQ_{1,1}^*$.

The other summand in the $\sQ_{1,2} \, \fprod \, \sQ_{1,2}$
should have $L_0$ eigenvalues in $\frac{1}{3} + \bN$. It is known in
general \cite{BBK-multiple_SLEs, Kytola-local_mgales}
that the partition function of another double SLE,
$\wtil{Z} = (x-y)^{h_{1,3} - 2 h_{1,2}} = Z_{\rho=2}$,
provides a local martingale
$\wtil{Z} / Z$ for SLE${}_\kappa(\kappa-6)$. In our case $\kappa = 6$
this means $\wtil{Z} = (x-y)^{1/3}$. Indeed
$A_{\rho=0} \, \wtil{Z} = 0$ and we compute
\begin{align*}
L_1 \, \wtil{Z} = 0 \qquad L_2 \, \wtil{Z} = 0 \qquad &
(L_0 - \frac{1}{3}) \, \wtil{Z} = 0
\end{align*}
to see that $\sU \, \wtil{Z} \subset \Kern A_{\rho=0}$ is a
highest weight module of weight $h = \frac{1}{3} = h_{1,3}$. To check
its irreducibility one verifies
\begin{align*}
\big( L_{-1}^3 -\frac{8}{3} L_{-2} L_{-1} + \frac{4}{9} L_{-3} \big)
    \; \wtil{Z} \; = \; & 0
\textrm{ .}
\end{align*}
This is enough since weight $h_{1,3}$ corresponds to a chain type Verma
module, whose maximal proper submodule is generated by the
grade $3$ singular vector.

\subsubsection{Comparison of the fusion and local martingales}
We may conclude that $\Theta$ and $\wtil{Z}$ generate the local
martingales that one expects from the fusion
$\sQ_{1,2} \, \fprod \, \sQ_{1,2}$
\begin{align*}
\Kern A_{\rho=0} \; \supset \; \sU \; \Theta \, + \, \sU \; \wtil{Z}
    \, \isom \, \sQ_{1,1}^* \, \oplus \, \sL_{h=1/3}^*
    \, = \, (\sQ_{1,2} \, \fprod \, \sQ_{1,2})^*
\textrm{ .}
\end{align*}
However, it seems fair to ask why the leftmost inclusion is not
an equality. For example the local martingale $\Xi$
(and much of the module $\sU \, \Xi$ generated by it)
does not appear in the contragredient of the fusion product.
So in fact $\sU \, \Theta$ is naturally included in a larger
module $\sU \, \Theta + \sU \, \Xi \, \subset \, \Kern A$.
An analogous phenomenon happens with the other part of the direct sum.
One may check that $\Upsilon(x,y,a_2,a_3,\ldots) =
(x-y)^{1/3} (\frac{2}{21} x^3 + \frac{1}{7} x^2 y
 + \frac{3}{7} x y^2 - a_3)$ is in $\Kern A_{\rho=0}$. Furthermore,
\begin{align*}
L_1 \, \Upsilon = \big( -\frac{45}{7} \, L_{-1}^2 
    + \frac{39}{7} \, L_{-2} \big) \, \wtil{Z}
    \; \in \; \sU \, \wtil{Z}
\quad \textrm{ ,} \quad L_2 \, \Upsilon = 0 \\
(L_0 - \frac{10}{3}) \, \Upsilon = 0
\end{align*}
and $L_3 \, \Upsilon = \wtil{Z}$, so we have
$\sL_{h=1/3} \isom \sU \, \wtil{Z} \subset
\sU \, \Upsilon$. Since $\Upsilon \notin \sU \, \wtil{Z}$
(one checks that
$\dmn (\bC \, \Upsilon + \sU_3 \, \wtil{Z} ) = 3 > 2 =
\dmn \sU_3 \, \wtil{Z}$)
and $L_n \, \Upsilon \in \sU \, \wtil{Z}$ for $n>0$, the quotient
$(\sU \, \Upsilon) / (\sU \, \wtil{Z})$ is a highest
weight module of weight $h=\frac{10}{3} = h_{1,3}+3 = h_{1,6}$.
Thus also $\sU \, \wtil{Z}$ is naturally contained in a larger
module.

To understand why the space of local martingales, $\Kern A_{\rho=0}$,
is larger than the fusion that we attempted to study, we should take
a look back at the constructions \cite{Kytola-local_mgales}
which showed that $\Kern A_{\rho=0}$ is a Virasoro module.
One notices that there is no need to make any assumptions of the
field located at $y$ except that
it is a primary field of the correct weight. In particular the
null field condition that would be satisfied if the field was
$\psi$, corresponding to $\sQ_{1,2}$, is not used. We thus expect
$\Kern A_{\rho=0}$ to rather reflect the fusion
$\sQ_{1,2} \, \fprod \, \Verma_{h=0}$ than
$\sQ_{1,2} \, \fprod \, \sQ_{1,2}$.
\begin{figure}
\begin{center}
\includegraphics[width=0.75\textwidth]{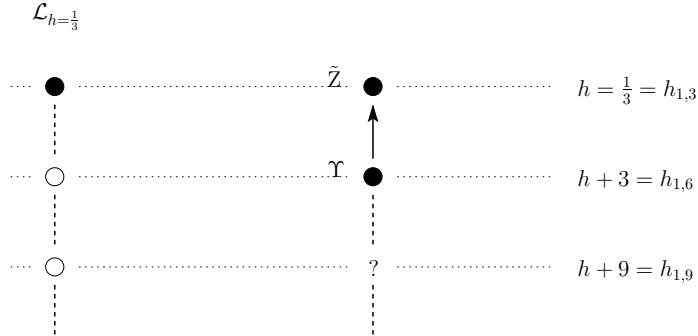}
\caption{\emph{The function $\wtil{Z}$
generates the
irreducible highest weight module $\sL_{h=1/3}^* \isom \sL_{h=1/3}$,
the structure of which is portrayed on the left.
It is however also contained in a larger module generated by $\Upsilon$
(portrayed on the right). The larger module is not contained
in the fusion $\sQ_{1,2} \, \fprod \, \sQ_{1,2}$ (see main text),
so we haven't examined the precise structure of it.
}}
\label{fig: L13 module}
\end{center}
\end{figure}

The constructions \cite{Kytola-local_mgales} do also provide a
probabilistic interpretation of guaranteeing the desired null vector
condition for the field at $y$. This consists of studying a double
SLE (a particular case of multiple SLEs \cite{BBK-multiple_SLEs})
with the same partition function $Z=(x-y)^{-2 h_{1,2}} = 1$.
The local martingales for this process form a smaller
module $\Kern A_{\rho=0} \, \cap \, \Kern \Arev_{\rho=0}$,
where $\Arev$ is in fact the generator of the chordal SLE in the
reverse direction\footnote{We point out
that the intersection $\Kern A_{\kappa ; \rho=\kappa-6} \, \cap \,
\Kern \Arev_{\kappa ; \rho=\kappa-6}$
has been used to obtain non-trivial evidence of the surprisingly
subtle question of the reversibility of chordal SLE
trace \cite{KK-reversibility_duality, Kytola-thesis}.},
from $y$ to $x$
\begin{align*}
\Arev_{\kappa ; \rho=\kappa-6} & \; = \;
    \frac{\kappa}{2} \; \frac{\partial^2}{\partial y^2}
    + \frac{2}{x - y} \; \pder{x}
    - \frac{2 \, h_{1,2}(\kappa)}{(x-y)^2} \\
& + 2 \sum_{l \geq 2} \Big( \sum_{\substack{0 \leq m \leq l-2 \\
                0 \leq r \leq \lfloor \frac{l-2-m}{2}\rfloor}} y^m 
    (-1)^r \frac{(m+r)!}{m! \; r!}
    \sum_{\substack{k_1, \ldots k_r \geq 2 \\ \sum k_j = l-m-2}}
    a_{k_1} \cdots a_{k_r} \Big) \; \pder{a_l}
\textrm{ .}
\end{align*}

Indeed, $\Arev \, Z = \Arev \, \Theta = 0$ but
$\Arev \, \Xi = 2 \, (x-y)^{-1} \neq 0$ and
$\Arev \, \Upsilon = \frac{20}{7} \, (x-y)^{4/3} \neq 0$. The
unexpected local martingales don't appear in the double SLE. We
conjecture that in general
\begin{align*}
\Kern A_{\kappa ; \rho=\kappa-6} \, \cap \,
\Kern \Arev_{\kappa ; \rho = \kappa-6} \; \overset{?}{=} \;
(\sQ_{\kappa ; 1,2} \, \fprod \, \sQ_{\kappa ; 1,2})^*
\textrm{ ,}
\end{align*}
since null field conditions of both fused parts are taken into
account in the double SLE local martingales.

\subsection{SLE${}_6(-2)$ and logarithms through fusion}
\label{sec: further fusion}

\subsubsection{The simplest fusion to produce logarithms in percolation}
The space of local martingales for a correctly chosen SLE variant
turned out to be appropriate for the fusion of two Cardy's fields
$\psi$. Of course requiring the fields to form a space that is
closed under fusion one should take that result, and further fuse
it with the other fields we have already found.
The next step in this direction is important as it will show that
the CFT of percolation necessarily contains logarithmic operators.
As in \cite{MR-percolation_LCFT} we next address the fusion
$\sQ_{1,2} \, \fprod \, \sL_{h=1/3}$, the result of which
is a staggered module $\sS$ with left
and right highest weight modules $\sQ_{1,2}$ and $\sQ_{1,4}$
respectively, i.e.
\begin{align*}
0 \longrightarrow \sQ_{1,2} \longrightarrow \sS
\longrightarrow \sQ_{1,4} \longrightarrow 0
\textrm{ .}
\end{align*}
The representation is characterized by giving the invariant $\beta$
defined in Section \ref{sec: Virasoro representations}.\footnote{For
reasons not very well understood before, in this case the invariant
turns out to have only one possible value, see
\cite{MR-logarithmic_M2p, KR-staggered}.}
From \cite{MR-percolation_LCFT} we quote the value $\beta = -\frac{1}{2}$.
For later comparison, the graded dimension is
\begin{align*}
\chi_{\sS} (q) \; = \; & \sum_{k \geq 0} (\dmn \sS_{[k]}) \; q^k
    \; = \; \sum_{k \geq 0}
        \big( \dmn (\sQ_{1,2})_k + \dmn (\sQ_{1,4})_{k-1} \big) \; q^k \\
= \; & 1 + 2 q + 2 q^2 + 4 q^3 + 6 q^4 + 8 q^5 + 12 q^6 + 17 q^7
      + 23 q^8 + \cdots
\textrm{ .}
\end{align*}

\subsubsection{An SLE variant that describes a crossing event}

We'd again like to find a probabilistic setup for studying the
fusion $\sQ_{1,2} \, \fprod \, \sL_{h=1/3}$. The starting
point of the SLE curve (or exploration path) provides the
field $\psi$ corresponding
to $\sQ_{1,2}$ and at another point we should have a primary field
of conformal weight $h=1/3$. In SLE${}_\kappa(\rho)$ the boundary
change at $Y_0$ is a primary with weight
$h(\rho) = \frac{\rho (\rho+4-\kappa)}{4\kappa}$,
%\cite{Kytola-SLE_kappa_rho},
so we should choose $\rho = -2$ or $\rho = 4$. Luckily, the case $\rho = -2$
has already been studied in Section \ref{sec: ell 1}
and even the probabilistic interpretation has been discussed in
Section \ref{sec: SLE variants}: we know that it describes an ordinary
chordal SLE$_{\kappa}$ from $X_0$ to $Y_0$ in $\bH$, conditioned to
reach the endpoint $Y_0$ without touching the interval $(Y_0,\infty)$.

In view of the discrete definition of the percolation exploration process,
the event on which we condition can be made more transparent.
The complementary event that the exploration path would reach $(Y_0,\infty)$
before $Y_0$ means the existence of a connected white cluster that joins
$(X_0,Y_0)$ to $(Y_0,\infty)$ --- indeed the hexagons that are immediately on
the right of the exploration path belong to such a cluster
(compare with Figure \ref{fig: exploration}(b)). The event on which we
condition should prevent the above kind of white cluster,
by having a black cluster
from exactly $Y_0$ (a single hexagon, say) to the interval $(-\infty,X_0)$.

In fact the discrete lattice
introduced a microscopic length scale $\delta$, the size of the single
hexagon. This suggests
that a naive equivalent in the continuum description is to require
crossing from $(-\infty,X_0)$ to $(Y_0^-,Y_0^+)$, where $Y_0^\pm \approx Y_0$
with $Y_0^+ - Y_0^- = \delta$. This is a crossing event \`a la Cardy,
and in the limit $\delta \rightarrow 0$ one brings two of the ``corners''
of the conformal quadrilateral together.
Conditioning SLE${}_6$ on Cardy's crossing event
is in fact easily done: the SLE variant\footnote{In fact it is very natural
to allow multiple SLEs in this context \cite{BBK-multiple_SLEs}.}
will have partition function given by Cardy's
formula, $Z(x, y_-, y_+) \, = \, \PR[\textrm{crossing}] \, = \,
F(\frac{y_- - x}{y_+ - x})$, where
\begin{align*}
F(r) \; = \; \frac{\Gamma(\frac{2}{3})}{\Gamma(\frac{1}{3})^2} \;
    \int_r^1 s^{-2/3} (1-s)^{-2/3} \; \ud s
\textrm{ .}
\end{align*}
Girsanov's formula gives the drift of the driving process under the
conditioned measure,
$\kappa \; \pder{x} \log Z(x,y_-,y_+)$ and in the limit
$y_-, y_+ \rightarrow y$ it tends to $\frac{-2}{x-y}$, i.e. the
conditioned process indeed tends to SLE${}_6(-2)$.
\begin{figure}
\begin{center}
\includegraphics[width=0.7\textwidth]{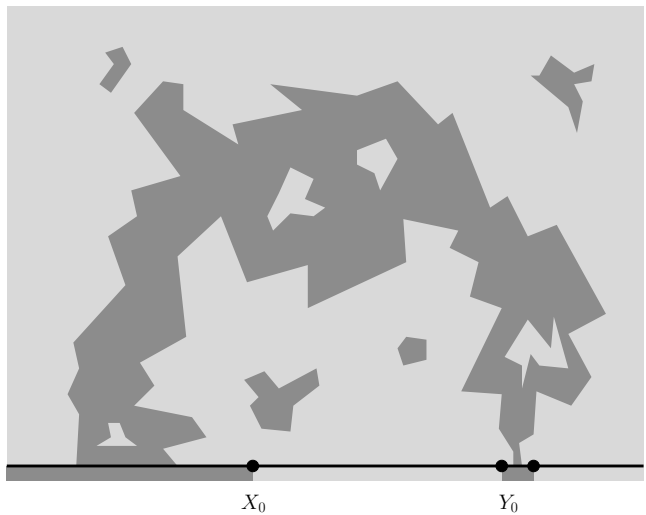}
\caption{\emph{The variant SLE${}_6(-2)$ can be obtained by
conditioning on a crossing event from a tiny interval around
$Y_0$ to $(-\infty, X_0)$. The picture clearly suggests that we're
bringing two Cardy's color changing boundary fields $\psi$ together
at this point. The conditioning picks the component $h=1/3$ instead
of $h=0$ from the fusion of two $\psi$ fields at $Y_0$.}}
\label{fig: subsequent fusion interpretation}
\end{center}
\end{figure}

\subsubsection{Local martingales for SLE${}_6 (-2)$}
For the module of local martingales $\Kern A_{\kappa=6;\rho=-2}$ we
can already quote many results from Section \ref{sec: ell 1}.
The partition function $Z_{\rho = -2} = (x-y)^{-1/3}$ generates a highest
weight module with $h=0$ and it has a nonzero singular vector at grade $1$.
The singular is in fact the partition function for the choice $\rho=4$,
$L_{-1} Z_{\rho=-2} = -\frac{1}{3} Z_{\rho=4}$, where
$Z_{\rho = 4} = (x-y)^{2/3}$.
The operator $A_{\rho=-2} \, = \, A_{\rho=4}$ is the same for the two
processes, while the local martingales of the two differ by a factor
$Z_{\rho=4}/Z_{\rho=-2} = (x-y)$ (recall that what we abusively called
the space of local martingales consists of functions that should still
be divided by the partition function to obtain actual local martingales).
It is now immediate that
$A_{\rho=-2} \; Z_{\rho=-2} \, = \, A_{\rho=-2} \; Z_{\rho=4}
\, = \, 0$. Observe also
that the formulas for $L_n$ are the same for these two SLE${}_6(\rho)$
processes.

We complete the analysis of Section \ref{sec: ell 1} in this degenerate
case at $\kappa=6$. To determine completely the structure of the highest
weight representation $\uea \, Z_{\rho=-2}$ we compute
\begin{align*}
\big( L_{-1}^2 - \frac{2}{3} L_{-2} \big) \, Z_{\rho=-2} = 0
\end{align*}
which confirms that
$\sU \, Z_{\rho = -2} \, \isom \, \sQ_{1,2}$. Without further
computations one concludes the irreducibility of
$\sU \, Z_{\rho = 4} \, \isom \, \sL_{h=1}$, although one
can also check directly the existence of null vectors at levels
$4$ and $6$.

The logarithmic partner to $L_{-1} Z_{\rho=-2}$ is $\Lambda$, and we have
%already
computed
\begin{align*}
(L_0 - 1) \; \Lambda \, = \, L_{-1} \; Z_{\rho=-2}
\quad \textrm{ and } \quad
L_1\; \Lambda \, = \, - \frac{1}{2} Z_{\rho=-2}
\textrm{ .}
\end{align*}
The latter already says that the invariant takes the same value as in the
corresponding fusion, $\beta = -1/2$, although we still haven't got enough
information to decide the structure of the right module.
By a computer assisted computation one obtains the numbers of linearly
independent local martingales by grade,
\begin{align*}
\chi_{\sU \, \Lambda} (q) \; = \; &
    \sum_{k \geq 0} \Big( \dmn \big( \sU^-_{k-1} \, \Lambda
    + \sU^-_{k} \, Z_{\rho=-2} \big) \Big) \; q^k \\
= \; & 1 + 2q + 2 q^2 + 4 q^3 + 6 q^4 + 8 q^5 + 12 q^6 + 17 q^7 + \cdots
\textrm{ ,}
\end{align*}
in agreement with the character of the module $\sS$.
Noting the graded dimensions of the left highest weight module
$\sU \, Z_{\rho=-2}$ (which can be computed directly, but also follows
from the earlier verification of $\uea Z_{\rho=-2} = \sQ_{1,2}$)
\begin{align*}
\chi_{\sU \, Z_{\rho=-2}} (q) \; = \; &
    \sum_{k \geq 0} \dmn \big( \sU^-_k \, Z_{\rho=-2} \big) \; q^k \\
= \; & 1 + 1 q + 1 q^2 + 2 q^3 + 3 q^4 + 4 q^5 + 6 q^6 + 8 q^7 + \cdots
\end{align*}
we deduce the graded dimension of the quotient
$(\sU \, \Lambda) / (\sU \, Z_{\rho=-2})$
\begin{align*}
& \sum_{k \geq 0} \Big( \dmn \big( \sU^-_k \, \Lambda
    + \sU^-_{k+1} \, Z_{\rho=-2} \big)
    \; - \; \dmn \sU^-_{k+1} \, Z_{\rho=-2} \Big) \; q^k \\
= \; & 1 + q + 2 q^2 + 3 q^3 + 4 q^4 + 6 q^5 + 9 q^6 + \cdots \\
= \; & p(0) + p(1) q + p(2) q^2 + p(3) q^3 + (p(4)-p(0)) q^4 \\
& \qquad + (p(5)-p(1)) q^5 + (p(6)-p(2)) q^6 + \cdots
\textrm{ .}
\end{align*}
This confirms that the quotient is isomorphic to $\sQ_{1,4}$:
at grade $4$ there is a null vector whereas
at grade $6$ there is a non-zero singular vector (corresponding
to a sub-singular vector at grade $7$ of the staggered module
$\uea \, \Lambda$).
\begin{figure}
\begin{center}
\includegraphics[width=1.0\textwidth]{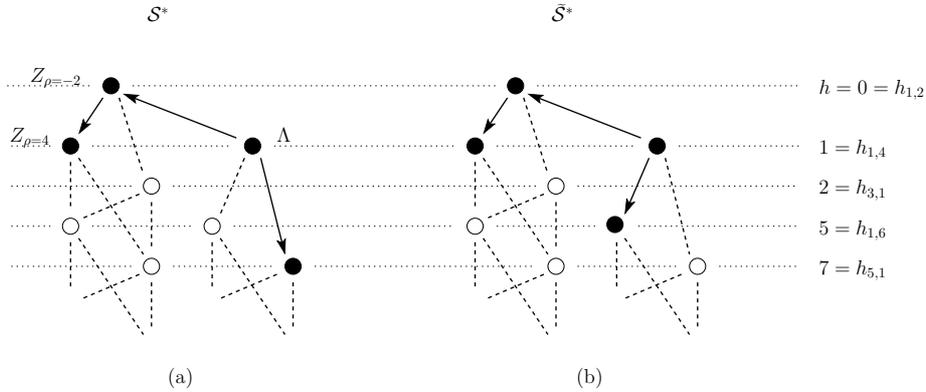}
\caption{(a) \emph{The local martingales for SLE${}_6(-2)$ contain a
staggered module $\sS^*$ contragredient to the fusion
$\sQ_{1,2} \, \fprod \, \sL_{h=1/3} \; = \; \sS$.
}
(b) \emph{A very similar staggered module of local martingales is found for
the SLE variants of Section \ref{sec: existence of Q21}. The existence of
such a module in the conformal field theory of percolation has been argued
to be impossible.}}
\label{fig: log module}
\end{center}
\end{figure}

Above, we have effectively shown that
$\uea \Lambda \subset \Kern A_{\kappa=6;\rho=-2}$
is isomorphic to the result of the fusion
$\sQ_{1,2} \, \fprod \, \sL_{h=1/3}$.
This is not exactly what we claimed, as we have not yet taken contragredients.
But just like in Section \ref{sec: contragredients}, elementary means
allow to check that the module is contragredient to itself. A small
difference is that we must now
show that $\uea \xi^*$ has a null vector at grade $2$,
which however follows rather directly from graded dimensions. Indeed,
we know that $\uea \xi^*$ and $\uea \eta^* / \uea \xi^*$ are highest
weight modules of highest weights $0$ and $1$ respectively, so the inequality
$2 = \dmn (\uea \Lambda)_{[2]} \geq
\dmn (\uea \eta^* / \uea \xi^*)_{[2]} + \dmn (\uea \xi^*)_{[2]} =
1 + \dmn (\uea \xi^*)_{[2]}$ leaves no other possibility but to have a null
vector $(L_{-1}^2 - \frac{2}{3}L_{-2}) \xi^* = 0$ (the grade one 
singular $L_{-1} \xi^*$ is non-zero by earlier analysis).
Similarly, graded dimensions also imply that $\uea \eta^* / \uea \xi^*$ must
have a null vector at grade $4$. The last remaining check of
non-zero singular vector at grade $6$ can be done mechanically\footnote{For
completeness we indicate one way of checking this last property. The
singular vector at grade $6$ of the Verma module $\Verma_{h=1}$ takes the form
$\chi_{3,2} \, v_{h=1}$, where the formula for $\chi_{3,2} \in \uea^-_6$
was given in Section \ref{sec: Cardys field}.
We note a peculiarity of $c=0$, $h=0$, which allows us to write any
$U \in \uea^-_m$ for $m \geq 2$ as $U = U' \chi_{1,1} + U'' \chi_{1,2}$, where
$\chi_{1,1}=L_{-1}$ and $\chi_{1,2}=L_{-1}^2-\frac{2}{3} L_{-2}$ are singulars.
Therefore any grade $7$ vector in $\uea \xi^*$ annihilates
$\chi_{3,2} \Lambda$, as one has for any $U \in \uea^-_{7}$
$$\bra U \xi^* , \chi_{3,2} \Lambda \ket =
\bra \chi_{3,2}^\dagger (U' \chi_{1,1} + U'' \chi_{1,2}) \xi^*,
\Lambda \ket = \bra 0, \Lambda \ket = 0$$
(the latter term is zero by
consideration of grades while the first is zero because $\chi_{3,2}$ is
singular and $\chi_{1,1} \xi^*$ generates a highest weight module of weight
$h=1$).
To conclude that there is a non-zero singular vector at grade $6$ in the
highest weight module $\uea \eta^* / \uea \xi^*$ it is therefore enough
to show that $\chi_{3,2} \eta^*$ has a non-zero value on $\chi_{3,2} \Lambda$.
With the aid of a computer we indeed straightforwardly compute
\begin{align*}
\bra \chi_{3,2} \eta^* , \chi_{3,2} \Lambda \ket = &
    \bra \chi_{3,2}^\dagger \chi_{3,2} \eta^* , \Lambda \ket =
    \bra \Big(\frac{17248000}{243}-\frac{17248000}{81}\beta \Big) L_{-1} \xi^*
            , \Lambda \ket = - \frac{21 560 000}{243} \neq 0
\textrm{ .}
\end{align*}}
We then conclude that the contragredient to the appropriate fusion product
has been realized as local martingales again,
$$ (\sQ_{1,2} \, \fprod \, \sL_{h=1/3})^* \isom
\sU \, \Lambda \, \subset \, \Kern A_{\rho=-2}
\textrm{ .}$$

\subsection{On the question of existence of $\sQ_{2,1}$}
\label{sec: existence of Q21}

In \cite{MR-percolation_LCFT} it is argued that a consistent chiral CFT
with $c=0$ can't contain both $\sQ_{1,2}$ (the Cardy's field)
and $\sQ_{2,1}$. The question of whether $\sQ_{2,1}$ is present in percolation
CFT has nevertheless been subject to some discussion, as for example building
modular invariant partition function seems to require the character
of $\sQ_{2,1}$ \cite{Flohr-private_comm}. The percolation literature also
contains some references to fields of dimension
$h=h_{2,1}=\frac{5}{8}$, e.g. \cite{CZ-integrable_field_theory_of_Potts}.
In the SLE context a natural way for $\sQ_{2,1}$
to appear would be through Duplantier duality
$\kappa \leftrightarrow \frac{16}{\kappa} = \kappa^*$, see
\cite{Duplantier-conformally_invariant_fractals,
Zhan-duality_of_chordal_SLE, Zhan-duality_of_chordal_SLE_II,
Dubedat-duality_of_SLEs}, although this doesn't directly give an
interpretation of it as a boundary condition changing field for percolation,
either. Below we will outline a possible way to gain some
understanding of this question from the point of view of the local
martingales of SLE variants involving both $\kappa=6$ and the dual value
$\kappa^*=\frac{8}{3}$.

\subsubsection{The conflicting module}

The inconsistency of adding $\sQ_{2,1}$ to the theory ultimately boils down
to the observation that it would lead, through fusions, to the existence of
staggered modules with logarithmic couplings different from what
is produced by the fusions generated by $\sQ_{1,2}$ alone. This is argued
to contradict requirements of chiral conformal field theory
\cite{MR-logarithmic_M2p}.
From SLE point of view perhaps the simplest such conflicting module arises
in the fusion
$\sQ_{1,2} \; \fprod \; \sQ_{2,1} \; \fprod \; \sQ_{2,1}$.
From the computations in \cite{MR-logarithmic_M2p} and associativity of
fusion product one finds that this fusion contains a staggered module
$\wtil{\sS}$ characterized by the
exact sequence $0 \rightarrow \sQ_{1,2} \rightarrow
\wtil{\sS} \rightarrow \sQ_{3,2} \rightarrow 0$ and logarithic coupling
$\wtil{\beta} = 1/3$, i.e.
$L_1 \, \eta = \frac{1}{3} \, \xi$.
The structure of this staggered module is very similar to that of $\sS$
in Section \ref{sec: further fusion}, see Figure \ref{fig: log module},
but the logarithmic couplings are different $\beta \neq \wtil{\beta}$.

There are several possibile SLE${}_6$ variants that reflect the fusion
$\sQ_{1,2} \, \fprod \, \sQ_{2,1} \, \fprod \, \sQ_{2,1}$,
for example\footnote{The local martingales of the different choices
differ by a factor that is a ratio of partition functions, the structure
of the module of local martingales doesn't depend on this choice.
As we now have two passive points $y_1, y_2$, one for each module
$\sQ_{2,1}$, we need SLE variants slightly more general than in previous
sections. The interested reader will find details in
\cite{Kytola-local_mgales}.} SLE${}_{\kappa=6}(\rho_1=-3,\rho_2=-3)$
with partition function
\begin{align*}
Z_{\underline{\rho}=(-3,-3)} \; = \;
    \frac{(y_1-y_2)^{3/4}}{(x-y_1)^{1/2} \; (x-y_2)^{1/2}}
\textrm{ .}
\end{align*}
This is a highest weight vector of $L_0$ eigenvalue $1=h_{3,2}=h_{1,4}$.

In $\Kern A_{\kappa=6; \underline{\rho}=(-3,-3)}$ we find also another
highest weight vector
\begin{align*}
F \; = \; \frac{y_1 + y_2 - 2x}
    {(x-y_1)^{1/2} \; (x-y_2)^{1/2} \; (y_1-y_2)^{5/4}}
\textrm{ ,}
\end{align*}
which could serve as a partition function of an SLE variant as well.
The highest weight is $0=h_{1,1}=h_{1,2}$ and the level $1$ singular vector
is non-vanishing and in fact proportional to the above partition function
$L_{-1} \, F = \frac{1}{2} \, Z_{\underline{\rho}=(-3,-3)}$. We have
$(L_{-1}^2 - \frac{2}{3} L_{-2}) \, F = 0$ so that
$\sU \, F \isom \sQ_{1,2}$. Since we're looking for $\wtil{\sS}^*$,
the task now is to find the logarithmic partner $\Lambda$ of
$Z_{\underline{\rho}=(-3,-3)}$.

We note that $F$ and $Z_{\underline{\rho}=(-3,-3)}$ are annihilated
by $A_{\kappa=6; \underline{\rho}=(-3,-3)}$, as they should. But moreover
they satisfy constraints of the null vectors in $\sQ_{2,1}$:
they are annihilated also by the generators
$A^{(y_j)}_{\kappa=8/3; \underline{\rho}=(2,-4/3)}$ ($j=1,2$) of
SLE${}_{8/3}$ variants started from $y_1$ and $y_2$ with the same
partition function.
Also the logarithmic partner becomes determined
(up to the usual freedom in the choice of $\eta$)
if we require that it is in
$\Kern A_{\kappa=6; \underline{\rho}=(-3,-3)} \; \cap \;
\Kern A^{(y_j)}_{\kappa=8/3; \underline{\rho}=(2,-4/3)}$
(either one of $j$ equals $1$ or $2$ is enough, the other then follows).
We can choose
\begin{align*}
\Lambda \; = \; \frac{ 2 \, x \, (y_1 + y_2 - 2x) \;
          + \; 3 (y_1-y_2)^2 \, \log (y_1-y_2)}
    {6 \, (x-y_1)^{1/2} \; (x-y_2)^{1/2} \; (y_1-y_2)^{5/4}}
\textrm{ .}
\end{align*}
This choice satisfies the normalization
$(L_0 - 1) \, \Lambda = L_{-1} \, F$ and the logarithmic coupling is
$\wtil{\beta} = \frac{1}{3}$ as follows from $L_1 \, \Lambda = \frac{1}{3} F$.

It turned out to be computationally slightly too demanding to verify that
the quotient $(\sU \, \Lambda) / (\sU \, F)$ possesses a non zero singular
vector at level $4$ and a null vector at level $6$. But accepting that,
one would conclude that $\sU \, \Lambda$ is isomorphic to the contragredient
of the staggered module $\wtil{\sS}$ which appears in the fusion 
$\sQ_{1,2} \, \fprod \, \sQ_{2,1} \, \fprod \, \sQ_{2,1}$.

\subsubsection{Interpreting the variants in percolation}

Some experts of the conformal field theory approach to percolation
have found the above discussion of the conflicting module through
multiple SLEs confusing.
What we have shown so far is that there are multiple SLE
variants involving SLE type curves with both\footnote{These values
of $\kappa$ are the only ones consistent with $c=0$. Also in general,
the only consistent possibilities for multiple SLEs are such that only the
two dual values $\kappa$ or $\kappa^*=\frac{16}{\kappa}$
are used for different curves, see
\cite{Dubedat-commutation, Graham-multiple_SLEs, Kytola-local_mgales}.}
$\kappa=6$ and
$\kappa=\frac{8}{3}$ whose local martingales contain the staggered module
$\wtil{\sS}$
(one can choose as partition function any linear combination of $F$ and
$Z_{\kappa=6;\underline{\rho}=(-3,-3)}$) ---
and nothing yet about its relation to percolation.
Now in order to be specific, we will present a definite
conjecture that would relate the above multiple
SLE variant with partition function $F$ to a reasonable question about
triangular lattice site percolation.
Of course in doing so we risk being incorrect
at several places --- more careful research may then either clarify
and correct the claims made, or show them altogether wrong.

We start with comments on the roles on boundary conditions of $\sQ_{1,2}$
and $\sQ_{2,1}$. The interpretation of $\sQ_{1,2}$ (Cardy's field)
has become quite clear by now:
it corresponds to the starting of an exploration path from boundary
point at which the boundary colors change. If we
were to choose the variant with partition function $F$, the
field at infinity would also have this weight $h=h_{1,2}=0$.
To interpret $\sQ_{2,1}$ we turn to a duality for conformally
invariant random fractals that was observed by Duplantier
\cite{Duplantier-conformally_invariant_fractals}.
Precise proofs in the SLE context have been given in
\cite{Zhan-duality_of_chordal_SLE, Zhan-duality_of_chordal_SLE_II,
Dubedat-duality_of_SLEs}, and they relate SLE${}_\kappa$
to SLE${}_{\kappa^*}$, where $\kappa^* = \frac{16}{\kappa}$.
Roughly speaking the relation is that for $\kappa>4$,
the boundary of the region surrounded by SLE${}_\kappa$ trace looks locally
like SLE${}_{\kappa^*}$. Therefore, since
$h_{2,1}(\kappa) = h_{1,2}(\kappa^*)$ is the conformal weight of the
operator at the starting point of SLE${}_{\kappa^*}$,
the boundary changing fields
$\sQ_{2,1}$ may correspond to points at which an external perimeter
(this should be the correct percolation analogue of the boundary
of the region surrounded by the exploration path
\cite{ADA-percolation_external_perimeter})
of a percolation cluster reaches the boundary.

Guided by the above discussion let us imagine the
following situation in triangular lattice site percolation in
$D^\delta = \bH \cap \delta T$. The boundary hexagons are chosen
black on $(-\infty,x^\delta) \cup (y_1^\delta,y_2^\delta)$,
and white on $(x^\delta,y_1^\delta) \cup (y_2^\delta,\infty)$,
where $x^\delta < y_1^\delta < y_2^\delta$. We can define the external
perimeter of the white cluster (for example)
as the simple lattice path on white hexagons that starts from
$(x^\delta,y_1^\delta)$ and ends on $(y_2^\delta,\infty)$ and surrounds
the minimal number of hexagons (so it follows as closely as possible
the finite black cluster, but due to the requirements that it is simple
and uses only white sites, it can't enter the fjords). Now condition
on the event that this external perimeter starts precisely at the
hexagon on the left of $y_1^\delta$ and ends precisely at the hexagon on
the right of $y_2^\delta$.

In the above situation one can start tracing three random
curves from the boundary: the exploration path from $x^\delta$, or the
external perimeter from either of its two ends $y_1^\delta$ or $y_2^\delta$.
Now there are two important questions:
does the joint law of these curves have a scaling
limit as $\delta \searrow 0$ and
$(x^\delta,y_1^\delta,y_2^\delta) \rightarrow (x,y_1,y_2)$,
and if it does, is it conformally invariant. Supposing
the answer to both is affirmative,
one would expect the exploration path still to
look locally like an SLE${}_{\kappa=6}$, while a segment from
either end of the external perimeter should look like
SLE${}_{\kappa=8/3}$. Furthermore, the curves are to be built on the same
probability space, so multiple SLEs \cite{BBK-multiple_SLEs,
Graham-multiple_SLEs} (or a commuting SLE in the terminology of
\cite{Dubedat-commutation}) seem like plausible candidates for the
joint law. There are two linearly independent possibilities for
partition functions of such variants of multiple SLEs:
$Z_{\kappa=6;\underline{\rho}=(-3,-3)}$ and $F$.
The asymptotics of these as $|y_2 - y_1| \rightarrow 0$ are different.
By methods similar to \cite{BBK-multiple_SLEs} one can then argue
that only $F$ corresponds to a growth process in which the
$\kappa^*=\frac{8}{3}$ curves from $y_1$ and $y_2$ will meet, as they
should if they are the two ends of the external perimeter of a black
cluster that doesn't reach to infinity.

To summarize, the conjecture is that if we condition percolation on the
event that the external perimeter of a black cluster attached to
$(y_1,y_2)$ extends precisely from $y_1$ to $y_2$, then the scaling
limit of the three curves is the multiple SLE with partition function $F$.
Although there are situations where discrete interfaces 
have been proved to have scaling limit described
by SLE curves, the present conjecture (even if correct) is probably
too complicated for rigorous methods.
It would nevertheless be very
interesting to gain any probabilistic understanding of the case,
and to better understand the
relation to the inconsistency argument that would ban $\wtil{\sS}$
from the conformal field theory of percolation.

\subsection{On the operator content of percolation}

Above we considered the simplest but arguably the most fundamental fusions
in the conformal field theory of percolation along the lines of the
recent study \cite{MR-percolation_LCFT}. For each of these cases,
an SLE variant with a clear interpretation in terms of percolation
was found, and local martingales of that variant coincided with the
contragredient of the fusion product. This can be thought of as a confirmation
of the results, and as a direct link from the discrete percolation
model to the operator content question. However, we also presented
a conjecture that some situations in percolation are described by
the multiple SLEs whose module of local martingales has a structure
conflicting consistent chiral CFT.

We remark that in the quest for better understanding the
conformal field theory of
percolation there are still issues to be solved, notably
modular invariance poses a challenge in logarithmic conformal
field theories \cite{FGST-logarithmic_characters_and_modular_transformations,
FG-logarithmic_torus_amplitudes, Flohr-private_comm}. Whether the present
approach will be useful to resolve these or other
questions remains to be seen,
but the present note nevertheless seems to provide a new perspective to
the topic.

%  End Of Section  *************************************************
%  *****************************************************************

%  *****************************************************************
%  *  CONCLUSIONS  *************************************************
%  *****************************************************************
\section{Discussion and conclusions}
\label{sec: conclusions}

In this note we studied the local martingales of the SLE${}_\kappa$ growth
processes. The space of local martingales forms a Virasoro module,
whose structure we examined in several cases. We were particularly
interested in a phenomenon encountered in logarithmic conformal field theory,
namely non-diagonalizability of $L_0$. It was observed that this takes place
in natural SLE variants, and at all values of $\kappa>0$ and therefore at
all central charge $c \leq 1$. As a byproduct, we obtained values of
logarithmic couplings (invariants of the representations) by easy
calculations. These values generalized conjectured infinite series of
logarithmic couplings for logarithmic M$(2,p)$ minimal models.
The calculations in our case were rather harmless,
so there is a prospect that similar methods might provide easy access to
the more general conjectures made in \cite{MR-logarithmic_M2p}.

In all the considered cases, a close relation between the local martingales
and the fusion product of boundary condition changing fields was found
(the fusion products have been computed in the CFT literature
\cite{Gaberdiel-fusion_in_CFT, GK-indecomposable_fusion_products,
EF-fusion_for_augmented_minimal_models, RS-associative_algebraic_LCFT,
RP-fusion_algebra_of_percolation, MR-percolation_LCFT, MR-logarithmic_M2p}).
The contragredient of the fusion product module could be realized as local
martingales. We conjecture that this will hold in general, i.e.
for any SLE variant. In fact the local martingale condition is closely
related to a null descendant of one of the boundary fields --- for fusions
of several fields with null descendant, appropriate intersections of
modules should be taken, as we illustrated with an example
of double SLE. It would be interesting to see if the relationship
may lend to alternative definitions of fusion, or at least to efficient
computational methods to study fusions.

We considered in some detail the case $\kappa=6$, for which SLE describes the
scaling limit of critical percolation exploration path.
We compared the results to the recent computations of fusion algebra of
percolation \cite{MR-percolation_LCFT, RP-fusion_algebra_of_percolation,
EF-fusion_for_augmented_minimal_models, RS-associative_algebraic_LCFT},
and found agreement in the above
mentioned sense. In the cases of the most fundamental fusions, we exhibited
probabilistic interpretations of the SLE variants used. For example,
logarithms first arise in a SLE variant that is obtained by
conditioning on a crossing event. We also exhibited SLE variants whose
local martingales form a staggered module whose existence in the CFT of
percolation has previously been argued impossible, and discussed whether
this variant could nevertheless describe some situations in percolation.

The main philosophical message of the note is the relation of SLEs to
logarithmic conformal field theories. Indications of this sort of relation
have been observed in different ways in
\cite{SPR-geometric_exponents, CS-twist_operator_correlations}, and 
in fact it has been suggested that the CFT
description of non-local questions (such as those of the SLE curve)
must typically be logarithmic. It is reassuring that the different
approaches point to a similar conclusion, and having these different
perspectives should be useful when the relation of SLE and LCFT will be
better studied in the future.

%  End Of Section  *************************************************
%  *****************************************************************

\bigskip
\bigskip

{\bf Acknowledgements:}
First of all I wish to thank Michel Bauer and David Ridout for numerous
interesting discussions at all stages of the research.
I am grateful to Annekathrin M\"uller-Lohmann for explanations on
logarithmic conformal field theory and to Hans Plesner Jakobsen
for pointing out a misconception of mine about representation theory. 
I thank Aapo Kyr\"ol\"a for his patience in giving programming
related advice. Michel Bauer, David Ridout and Michael Flohr made
valuable comments about the first version of the manuscript.

This work was partly done at the Laboratoire de Physique Th\'eorique et
Mod\`eles Statistiques (Universit\'e Paris Sud 11)
and partly at Section de Math\'ematiques (Universit\'e de Gen\`eve)
and supported by ENRAGE European Network MRTN-CT-2004-5616,
Swiss National Science Foundation and European Research Council.

\def\cprime{$'$} \def\cprime{$'$} \def\cprime{$'$}

\end{document}